\documentclass[pre,tighten]{revtex4}

\usepackage{amssymb,amsmath}

\usepackage{graphicx}

\begin{document}

\title{Rheology of granular mixtures under uniform
       shear flow: Enskog kinetic theory versus molecular dynamics simulations}
\author{Jos\'e Mar\'{\i}a Montanero}
\email[E-mail: ]{jmm@unex.es}
\affiliation{Departamento de Electr\'onica e Ingenier\'{\i}a Electromec\'anica,
Universidad de Extremadura, E-06071 Badajoz, Spain}
\author{Vicente Garz\'{o}}
\email[E-mail: ]{vicenteg@unex.es}
\affiliation{Departamento de F\'{\i}sica, Universidad de Extremadura, E-06071
Badajoz, Spain}
\author{Meheboob Alam}
\email[E-mail: ]{meheboob@jncasr.ac.in}
\affiliation{Engineering Mechanics Unit, Jawaharlal Nehru Center,
Jakkur P.O., Bangalore 560064,
India}
\author{Stefan Luding}
\email[E-mail: ]{s.luding@tnw.tudelft.nl}
\affiliation{Particle Technology, DelftChemTech,
Julianalaan 136, 2628 BL Delft, The Netherlands}

\begin{abstract}
The rheological properties for dilute and moderately dense granular
binary mixtures of smooth, inelastic hard disks/spheres under uniform shear
flow in steady state conditions are reported.
The results are based on the Enskog kinetic theory, numerically solved by
a dense gas extension of the Direct Simulation Monte Carlo
method for dilute gases.  These results are confronted to the
ones also obtained by performing molecular dynamics simulations
with very {good agreement for the lower densities and higher
coefficients of restitution}.  It appears
that the range of densities for which the Enskog equation applies
decreases with increasing dissipation.
\end{abstract}

\date{\today}
\maketitle

\section{Introduction}
\label{sec1}

Rapid granular flows can be well modeled by an idealized system of
smooth hard disks/spheres with inelastic collisions \cite{E00,PL01}.
Despite the simplicity of the model, it has been widely studied in the
last few years as a prototype to gain some insight into the ``microscopic''
understanding of physical mechanisms involved in granular flows.
One well established tool is the kinetic theory and, more specifically,
the Boltzmann equation for a low density gas \cite{GS95,D01,vNE01,BC01}.

In recent years,
some methods have been developed to solve it and get accurate
predictions over a wide range of parameters. For
finite higher densities, the Enskog kinetic equation \cite{BDS97,GD99}
can be considered  as the extension of the Boltzmann equation. As
for elastic collisions, the inelastic Enskog equation takes into
account spatial correlations through the pair correlation function
but neglects the effect of correlations between the velocities of
two particles that are about to collide (molecular chaos assumption).
Nevertheless, some deviations from molecular chaos have been
observed in molecular dynamics (MD) simulations of granular
fluids in driven, sheared, or homogeneous states
\cite{LMM98,ML98,SM01,SPM01,PTNE01}. Being not significant for dilute
systems \cite{BM04}, the velocity correlations increase with the
density. {Although the presence of these correlations restricts
the range of validity of the Enskog equation, it can still be
considered as a good approximation at moderate densities
since, for instance, the Enskog kinetic theory \cite{GD99}
successfully models the hydrodynamic
profiles obtained in recent NMR experiments on a three-dimensional
vibrofluidized granular medium \cite{YHCMW02} or from
simulations for a two-dimensional system, even at very high
densities \cite{MPB03}.}

Most of the comparisons carried out between kinetic theory results,
molecular dynamics (MD) and especially hard sphere event-driven (ED)
simulations have been devoted to monodisperse systems, where all
the grains have the same mass and size. Needless to say, a real
granular system is usually characterized by some degrees of
polydispersity in mass and size, which can lead to
segregation in an otherwise homogeneous mixture.
As kinetic theory predicts for the homogeneous state
\cite{GD99a,LS01,ML98,MG02}, MD simulations \cite{DHGD02,HCS} show that the
kinetic temperatures for each species are clearly different and
present a complex dependence on the parameters of the problem
(dissipation, mass ratio, size ratio, composition, and density).
{Good agreement for the temperature ratio between Enskog theory and
simulations is found for moderate densities and not too strong dissipation
\cite{DHGD02}.}

In the case of shearing flow states, recent MD simulations have been
performed \cite{CH02,AL03} to analyze the (bulk) rheological properties of
granular mixtures under uniform shear flow (USF) conditions.
Numerical results for the shear stress and the granular energy
were compared with those derived from a kinetic theory valid in
the low dissipation limit \cite{WA99,AL02b}. The comparison between
theory and simulation shows an excellent agreement in the
quasielastic limit even at large size disparities. Given that this
theory \cite{WA99,AL02b} only applies for weak inelasticity, it
assumes energy equipartition and the granular temperature coincides
with the partial temperatures of each species. However, beyond the
low-dissipation limit, the failure of energy equipartition in
granular fluids has not only been predicted by numerical ``experiments''
\cite{ML98,MG02,DHGD02,HCS} but also was observed in real experiments \cite{exp}.

The goal of this paper is to determine the rheological properties of a
granular binary mixture under USF conditions in the context of the Enskog
kinetic theory, and to compare the theory to ED simulations.
The USF is characterized by uniform density and temperature
and a constant velocity profile. In contrast to classical, elastic
fluids, a steady state is reached in the system when the continuous
loss of energy due to inelastic collisions is compensated for by the
viscous (shear) heating.
As a consequence, for a given shear rate, the granular
temperature as well as the rheological properties are functions of
the coefficients of restitution.

Two special situations allow for simple analytical results.
First, in the low density limit
the corresponding Boltzmann equation can be approximately
solved by Grad's method \cite{MG02a}. This solution compares
quite well with Monte Carlo simulations \cite{MG02a,MG03,GM03bis},
even for strong dissipation. Second, for small shear rates
but finite densities, the Enskog equation can be solved
by the Chapman-Enskog method  \cite{CC70} and an explicit
expression for the Navier-Stokes shear viscosity coefficient
has been obtained in the leading Sonine approximation \cite{GM03}.
It must be remarked that the kinetic theory derived
in Ref.\ \cite{GM03} differs from the one previously obtained
in Refs.\ \cite{WA99} and \cite{AL02b} since the former
should apply for an arbitrary degree of dissipation and takes into
account the effect of temperature differences on momentum transport.
Beyond the above two special cases
(dilute gas and Navier-Stokes approximation), Lutsko \cite{L04} has
recently solved the Enskog equation for a moderately dense mixture under USF
by using a generalized moment method. Although this theory
compares quite well with MD simulations, it appears too complicated
to be useful for practical purposes here, since it involves many different
collision integrals that must be solved numerically.
Instead, as an alternative to the above theory \cite{L04},
we compute the rheological properties
from a {\em numerical} solution of the Enskog equation
by means of an extension of the well-known Direct Simulation
Monte Carlo (DSMC) method \cite{B94} to dense gases \cite{MS96,MLH97,HM00}.

Given that MD simulation avoids any assumption inherent
in the Enskog equation, the comparison between MD and Enskog results
determines the validity and limitations of the kinetic theory.
This is the main motivation of our study.
Since the parameter space over which the Enskog equation
is verified is quite large, the test of the kinetic theory
is quite stringent. The comparison carried out here shows
that the dependence of rheology on mechanical properties and
state conditions obtained from the Enskog kinetic theory presents good
agreement with MD simulations, except at high density and
strong dissipation. In addition, the agreement between kinetic
theory and simulation results is better than the one previously
reported  for the homogeneous cooling state \cite{DHGD02}.

It must be noted that our analysis is restricted to
the {\em uniform} shear flow without paying attention to
the possible formation of particle clusters (microstructure).
Both, kinetic theory and
DSMC results intrinsically assume homogeneity, whereas in MD some
degrees of inhomogeneity in density, temperature or shear rate
can evolve in a steady state \cite{AL03b}. This can be a possible source of
discrepancies between the Enskog results and MD simulations. However,
in the USF problem, due to the viscous heating, inhomogeneities
are weaker than for free cooling systems where clusters
continuously grow in size. This is perhaps the main reason for which
the results found here for a sheared (driven) fluid compare better with
MD than those obtained in
the undriven case \cite{DHGD02}.

The plan of the paper is as follows. In Sec.\ \ref{sec2}
we review the Enskog kinetic theory for USF in the case of
inelastic systems along with brief descriptions of the  Monte Carlo
and molecular dynamics simulation methods for the USF.
The comparison between the Enskog results and MD
simulations is made in Sec.\ \ref{sec3}.
We close the paper  in Sec.\ \ref{sec4} with  a brief discussion of
the results presented in this paper.

\section{Enskog kinetic theory and simulation methods}
\label{sec2}

We consider a binary mixture of smooth hard spheres ($d=3$) or
disks ($d=2$) of diameters $\sigma _{1}$ and $\sigma _{2}$,
and masses $m_{1}$ and $m_{2}$. The collisions between particles
of different species are characterized by three independent (constant)
coefficients of normal restitution $\alpha_{11}$, $\alpha_{22}$,
and $ \alpha_{12}=\alpha_{21}$, with values $0<\alpha_{ij}\leq 1$,
where $\alpha_{ij}$ refers to collisions between particles of
species $i$ and $j$. Let us assume that the mixture is under uniform shear flow (USF).
From a macroscopic point of view, the USF is characterized
by a linear velocity profile
\begin{equation}
\label{3.1}
{\bf u}=ay{\bf {\widehat{x}}},
\quad a=\frac{\partial u_x}{\partial y}=\text{const.} \, ,
\end{equation}
where $a$ is the (constant) shear rate and ${\bf u}$ refers to
the (common) flow velocity.
In addition, the partial densities of each species
$n_i$ and the granular
temperature $T$ are uniform, while the mass and heat fluxes
vanish by symmetry reasons. The rheological properties
of the system are given from the pressure tensor ${\sf P}$, which
is the only relevant flux of the problem. According to
the energy balance equation,
the temperature changes in time due to the competition
between two mechanisms: on the one hand, viscous (shear) heating
and, on the other hand, energy dissipation in collisions.
In the steady state, both mechanisms cancel each other
and the (steady) temperature is obtained by equating the
production term due to shear work with collisional dissipation:
\begin{equation}
\label{3.3}
aP_{xy} =- \frac{d}{2}  n T \zeta.
\end{equation}
Here, $P_{xy}$ denotes the $xy$ element of the pressure tensor,
$n=n_1+n_2$ is the total number density and
$\zeta$ is the cooling rate due to collisions among all the
species. This steady shear flow state is what we want to analyze here.

The balance equation (\ref{3.3}) shows the intrinsic connection
between the shear field and dissipation in the system. This is a peculiar
feature of granular fluids since there is an internal mechanism for which
the collisional cooling sets the strength of the velocity gradient.
This contrasts with the description of USF for elastic fluids where a
steady state is not possible unless an external thermostat is
introduced \cite{GS03}. Therefore,
for given values of the parameters of the mixture, in the
steady state the reduced shear rate
(which is the relevant nonequilibrium parameter of the problem)
$a^*\propto a/\sqrt{T}$
is a function of the coefficients of restitution
$\alpha_{ij}$ only \cite{SGD04}.
In particular,
the {\it quasielastic} limit ($\alpha\to 1$)
naturally implies the limit of {\it small} shear rates ($a^*\ll 1$) and vice versa.
Since the limit of small shear rates also implies
a continuum-level description
of the Enskog kinetic equation at the Navier-Stokes order,
we may conclude that in the USF the full nonlinear dependence of the shear viscosity
on the shear rate cannot be obtained from the Navier-Stokes approximation,
at least for finite dissipation \cite{SGD04}.
We shall come back to this point while discussing our results.

At a microscopic level, the USF is generated by
Lees-Edwards boundary conditions \cite{LE72,AT89}
which are simply periodic boundary conditions in the
local Lagrangian frame ${\bf V}={\bf v}-{\sf a}\cdot {\bf r}$
and ${\bf R}={\bf r}-{\sf a}\cdot {\bf r}t$. Here, ${\sf a}$ is
the tensor with elements $a_{k\ell}=a\delta_{k x}\delta_{\ell y}$.
In terms of the above variables, the one-particle velocity distribution functions
$f_i({\bf V})$ ($i=1,2$) are uniform \cite{DSBR86} and the
Enskog equation takes the form
\begin{equation}
aV_y\frac{\partial}{\partial V_x}
f_{i}=\sum_{j=1}^2J_{ij}^{\text{E}}\left[ {\bf V}|f_{i},f_{j}\right] \;,
\label{3.4}
\end{equation}
where the Enskog collision operator $J_{ij}^{\text{E}}\left[ {\bf V}|f_{i},f_{j}\right]$ reads \cite{GM03}
\begin{eqnarray}
\label{3.5}
J_{ij}^{\text{E}}\left[ {\bf V}_{1}|f_{i},f_{j}\right] &=&\sigma _{ij}^{d-1}\chi_{ij}\int d{\bf V}
_{2}\int d\widehat{\boldsymbol {\sigma }}\,\Theta (\widehat{{\boldsymbol {\sigma }}}
\cdot {\bf g})(\widehat{\boldsymbol {\sigma }}\cdot {\bf g})  \nonumber \\ &&\times \left[ \alpha _{ij}^{-2} f_i({\bf V}_1')f_j({\bf V}_2'+a\sigma_{ij}\widehat{\sigma}_y{\widehat{{\bf x}}})
-f_i( {\bf V}_1)f_j({\bf V}_2-a\sigma_{ij}\widehat{\sigma}_y{\widehat{{\bf x}}})\right].
\end{eqnarray}
Here, $\boldsymbol {\sigma}_{ij}=\sigma_{ij} \widehat{\boldsymbol {\sigma }}$, with $\sigma _{ij}=\left( \sigma _{i}+\sigma _{j}\right) /2$ and $\widehat{\boldsymbol {\sigma}}$ is a unit vector directed along the line of centers from the sphere of species $i$ to the sphere of species $j$ upon collision (i.e. at contact). In addition,  $\Theta $ is the Heaviside step function, and ${\bf g}={\bf V}_{1}-{\bf V}_{2}$. The
primes on the velocities denote the initial values $\{{\bf V}_{1}^{\prime},
{\bf V}_{2}^{\prime }\}$ that lead to $\{{\bf V}_{1},{\bf V}_{2}\}$
following a binary collision:
\begin{equation}
{\bf V}_{1}^{\prime }={\bf V}_{1}-\mu _{ji}\left( 1+\alpha _{ij}^{-1}\right)
(\widehat{{\boldsymbol {\sigma }}}\cdot {\bf g})\widehat{{\boldsymbol {\sigma }}}
,\quad {\bf V}_{2}^{\prime }={\bf V}_{2}+\mu _{ij}\left( 1+\alpha
_{ij}^{-1}\right) (\widehat{{\boldsymbol {\sigma }}}\cdot {\bf g})\widehat{
\boldsymbol {\sigma}} , \label{2.3}
\end{equation}
where $\mu _{ij}=m_{i}/\left( m_{i}+m_{j}\right)$.
In addition, we have taken into
account that the pair correlation function $\chi_{ij}$ is uniform
in the USF problem. The expression for the pressure tensor ${\sf P}$ contains both {\em kinetic} and {\em collisional} transfer contributions, i.e.,
\begin{equation}
\label{2.13.1}
{\sf P}={\sf P}^{\text{k}}+{\sf P}^{\text{c}}.
\end{equation}
The kinetic part is given by
\begin{equation}
{\sf P}^{\text{k}}=\sum_{i=1}^2\,\int d{\bf v}\,m_{i}{\bf V}{\bf V}\,f_{i}({\bf
v}) , \label{2.14}
\end{equation}
while the collisional part is \cite{GM03}
\begin{eqnarray}
\label{3.6}
{\sf P}^{\text{c}}&=&\frac{1}{2}\sum_{i=1}^2\sum_{j=1}^2\frac{m_im_j}{m_i+m_j}\chi_{ij}\sigma_{ij}^d
(1+\alpha_{ij})\int d{\bf V}_1\int d{\bf V}_2\int d\widehat{\boldsymbol {\sigma }}\,\Theta (\widehat{{\boldsymbol {\sigma}}}
\cdot {\bf g})(\widehat{\boldsymbol {\sigma }}\cdot {\bf g})^2  \nonumber\\
& & \times
\widehat{\boldsymbol {\sigma }}\widehat{\boldsymbol {\sigma }}
f_i\left({\bf V}_1+a\sigma_{ij}\widehat{\sigma}_y{\widehat{{\bf x}}}\right)f_j({\bf V}_2).
\end{eqnarray}
The rheological properties of the mixture are obtained from
the knowledge of the elements of the pressure tensor.
Apart from these rheological
properties, it is also interesting to get the temperature ratio
$T_1/T_2$ where the partial temperatures $T_i$ are defined as
\begin{equation}
\label{3.10bis}
T_i=\frac{m_i}{d n_i}\int\, d{\bf V} \,V^2 f_i({\bf V}).
\end{equation}
The deviation of $T_1/T_2$ from unity is a measure of the breakdown of the
energy equipartition \cite{ML98}.
Finally, the cooling rate $\zeta$ is given by \cite{GM03}
\begin{eqnarray}
\label{3.7}
\zeta&=&\frac{1}{2d}\frac{1}{nT}\sum_{i=1}^2\sum_{j=1}^2\frac{m_im_j}{m_i+m_j}\chi_{ij}\sigma_{ij}^{d-1}
(1-\alpha_{ij}^2)\int d{\bf V}_1\int d{\bf V}_2\int d\widehat{\boldsymbol {\sigma }}\,\Theta (\widehat{{\boldsymbol {\sigma}}}
\cdot {\bf g})(\widehat{\boldsymbol {\sigma }}\cdot {\bf g})^3  \nonumber\\
& & \times
f_i\left({\bf V}_1+a\sigma_{ij}\widehat{\sigma}_y{\widehat{{\bf x}}}\right)f_j({\bf V}_2).
\end{eqnarray}

\subsection{Enskog simulation Monte Carlo method for USF}
\label{sub2a}

Needless to say, it becomes prohibitively difficult to solve analytically
the Enskog equation (\ref{3.4}) for arbitrary dissipation. As said in the Introduction,
a recent approximate solution to (\ref{3.4}) has been proposed \cite{L04}. This solution,
which is based on a generalization of the Grad method, leads to a closed
set of equations for the elements of the pressure tensor. However,
these equations are too complicated to be useful as a practical tool in general.
As an alternative to the above analytical method, an extension of the Direct
Simulation Monte Carlo (DSMC)
method \cite{B94} to dense gases \cite{MS96,MLH97,HM00} has been used to
numerically solve (\ref{3.4}). This method is
usually referred to as the Enskog Simulation Monte Carlo (ESMC) method.
This procedure was devised to mimic the dynamics
involved in the Enskog collision term (still assuming
operator-splitting and molecular chaos),
and it has been compared
to hard sphere simulations in vibrated and freely cooling systems
\cite{MLH97,LMM98} and more previously has been used to analyze
rheological properties of monocomponent systems for both the
elastic \cite{MS96} and the inelastic \cite{MG02,MGSB99} case.
For a detailed description of the method applied to inelastic systems,
we refer to the reader
to Refs. \cite{MG02} and \cite{GM03}.

From the simulations, one evaluates the kinetic (\ref{2.14}) and
 collisional (\ref{3.6}) contributions to the pressure tensor,
\begin{equation}
\label{4.5}
{\sf P}^{\text{k}}=\sum_{i=1}^{2} \frac{m_i n_i}{N_i}\sum_{k=1}^{N_i} {\bf V}_k {\bf V}_k\; ,
\end{equation}
and
\begin{equation}
{\sf P}^{\text{c}}=\frac{n}{2N\Delta t}\sum_{k\ell}^{\dagger} \frac{m_im_j}{m_i+m_j} \sigma_{ij}(1+\alpha_{ij})
({\bf g}_{k\ell}\cdot \widehat{\boldsymbol \sigma}_{k\ell})\widehat{\boldsymbol \sigma}_{k\ell}
\widehat{\boldsymbol \sigma}_{k\ell}\; ,
\end{equation}
where $N_i$ is the number of ``simulated'' particles of species $i$,
$N=N_1+N_2$ is the total number, $\Delta t$ is the time step,
 and the dagger ($\dagger$) means that the summation is
restricted to the accepted collisions. In addition, $k$ denotes a particle
of species $i$ and $\ell$ a particle of species $j$. To improve the statistics,
the results are averaged over a number ${\cal N}$ of independent realizations
or replicas. In our simulations we have typically taken a total number of
particles $N=N_1+N_2=10^5$, a number of replicas ${\cal N}=10$, and a time
step $\Delta t=3\times 10^{-3} \lambda_{11}/V_{01}(0)$. Here,
$\lambda_{11}=(\sqrt{2}n_1 \widetilde{\sigma}_{11})^{-1}$  is the
mean free path for collisions 1--1 and $V_{01}^2(0)=2T(0)/m_1$ {where
$T(0)$ is the initial temperature. Here, $\widetilde{\sigma}_{ij}$
means the total cross-section for collisions of type $i$--$j$ \cite{note}}.
More technical details on the application
of the ESMC method to USF can be found in Ref.\ \cite{GM03}.

In the simulations the initial velocity distribution function is that of local
equilibrium. The ESMC simulation is prepared as follows:
after an initial transient period,
the system reaches a steady state where the values obtained for the reduced quantities,
see below Eqs.\ (\ref{3.8})--(\ref{3.10}), are independent
of the initial preparation of the system.
Their values do depend on the values of the coefficients of restitution, the packing fraction,
and the ratios of mass, concentrations and sizes.

\subsection{Event-driven simulation method for uniform shear flow}
\label{sec2b}

The molecular dynamics (MD) and event-driven (ED) simulation methods
are extensively discussed, e.g.\ in textbooks \cite{AT89}, for
elastic systems, and were also applied to sheared granular systems, see
Refs. \ \cite{AL03}, \cite{AL02b}, \cite{AL02}, \cite{AL03b}
and references therein.
Also the preparation-procedure and system size dependences are
discussed in Ref.\ \cite{AL03} with no striking insights, so that we
do not repeat these details here.

\subsection{Rheology}
\label{sub2c}

The rheological properties of the mixture are determined from the
nonzero elements of the pressure tensor.  Possibilities to non-dimensionalize
are  to scale either by the relevant small species contributions \cite{AL03}
or by the corresponding mean values \cite{AL02}.  Applying the former scheme, one
arrives at the reduced shear viscosity
\begin{equation}
\label{3.8}
\mu=\frac{|P_{xy}|}{\varrho_1 \sigma_1^2 a^2},
\end{equation}
the reduced temperature
\begin{equation}
\label{3.9}
\theta=\frac{T}{m_1 \sigma_1^2 a^2},
\end{equation}
and the reduced pressure
\begin{equation}
\label{3.10}
\Pi=\frac{p}{\varrho_1 \sigma_1^2 a^2}, \quad p=\frac{1}{d} {\text{tr}}\,{\sf P}.
\end{equation}
Here ${\text{tr}}\,{\sf P}$ denotes the trace of the pressure tensor ${\sf P}$
and $\varrho_i$ is the material density of species $i$. In the case of hard
disks $\varrho_i=4m_i/\pi \sigma_i^2$ while
$\varrho_i=6m_i/\pi \sigma_i^3$ for hard spheres.

\begin{figure}[htb]
\begin{center}
\begin{tabular}{lr}
\resizebox{6.2cm}{!}{\includegraphics{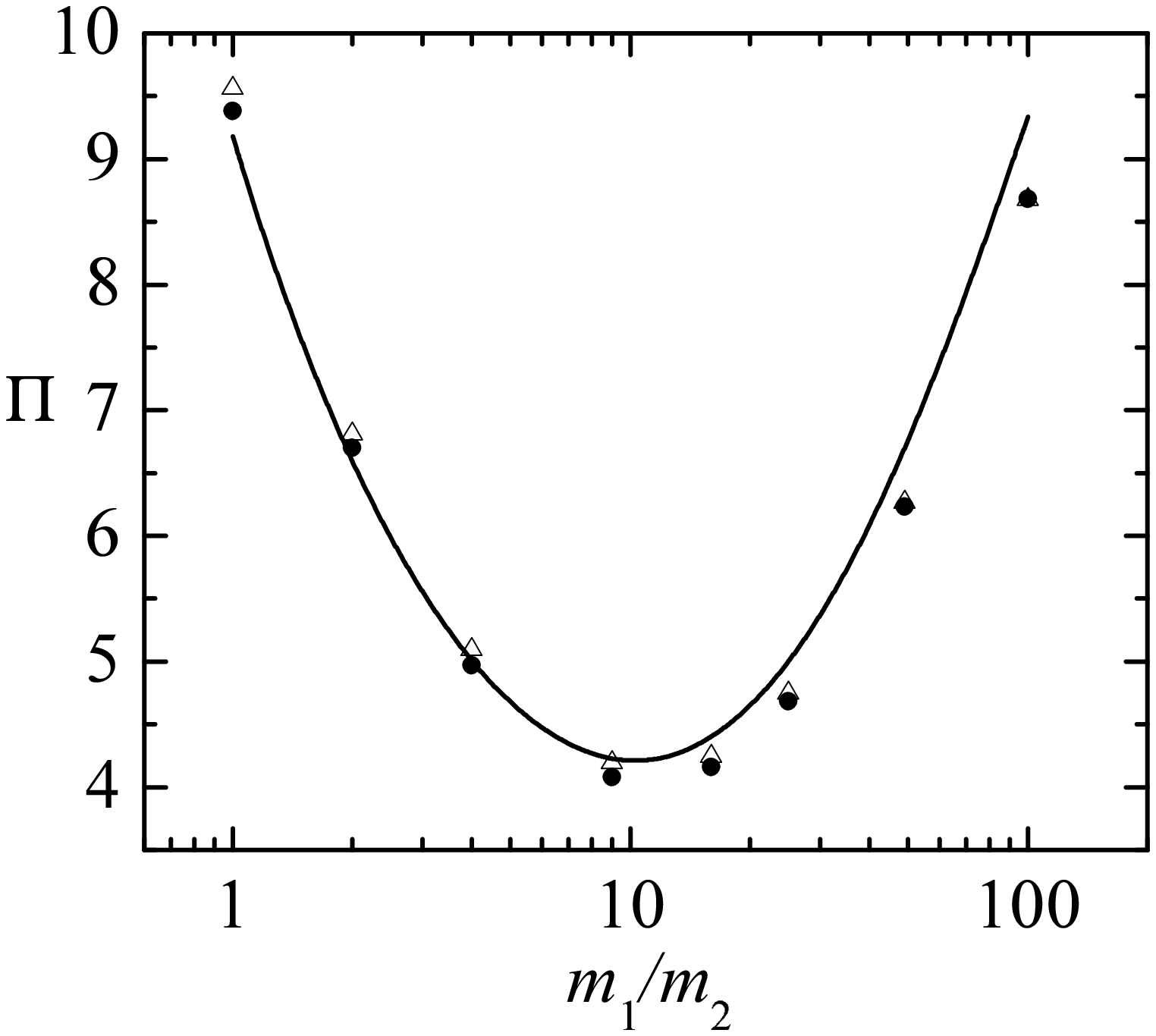}}&\resizebox{6.6cm}{!}{\includegraphics{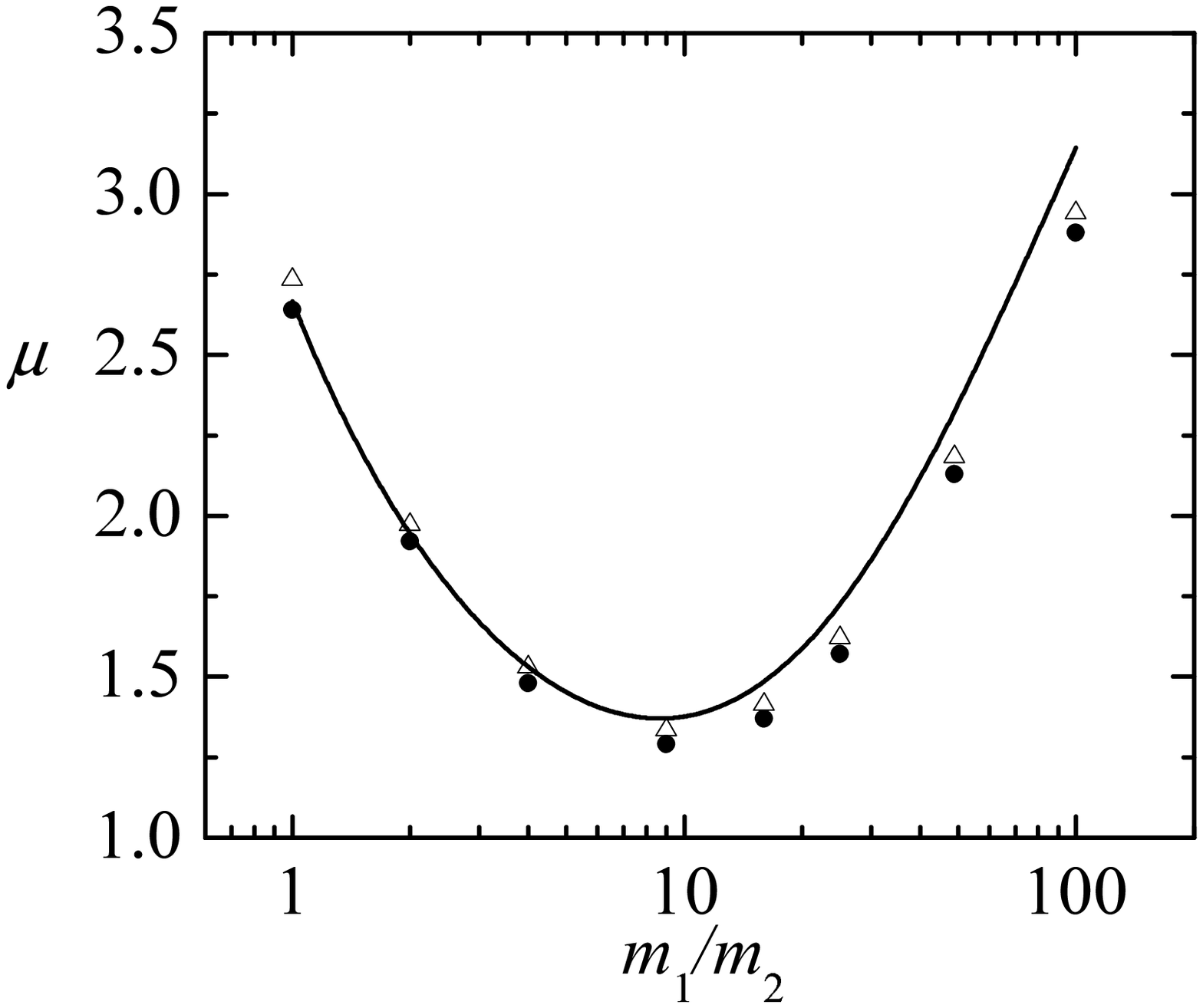}}\\
\resizebox{6.5cm}{!}{\includegraphics{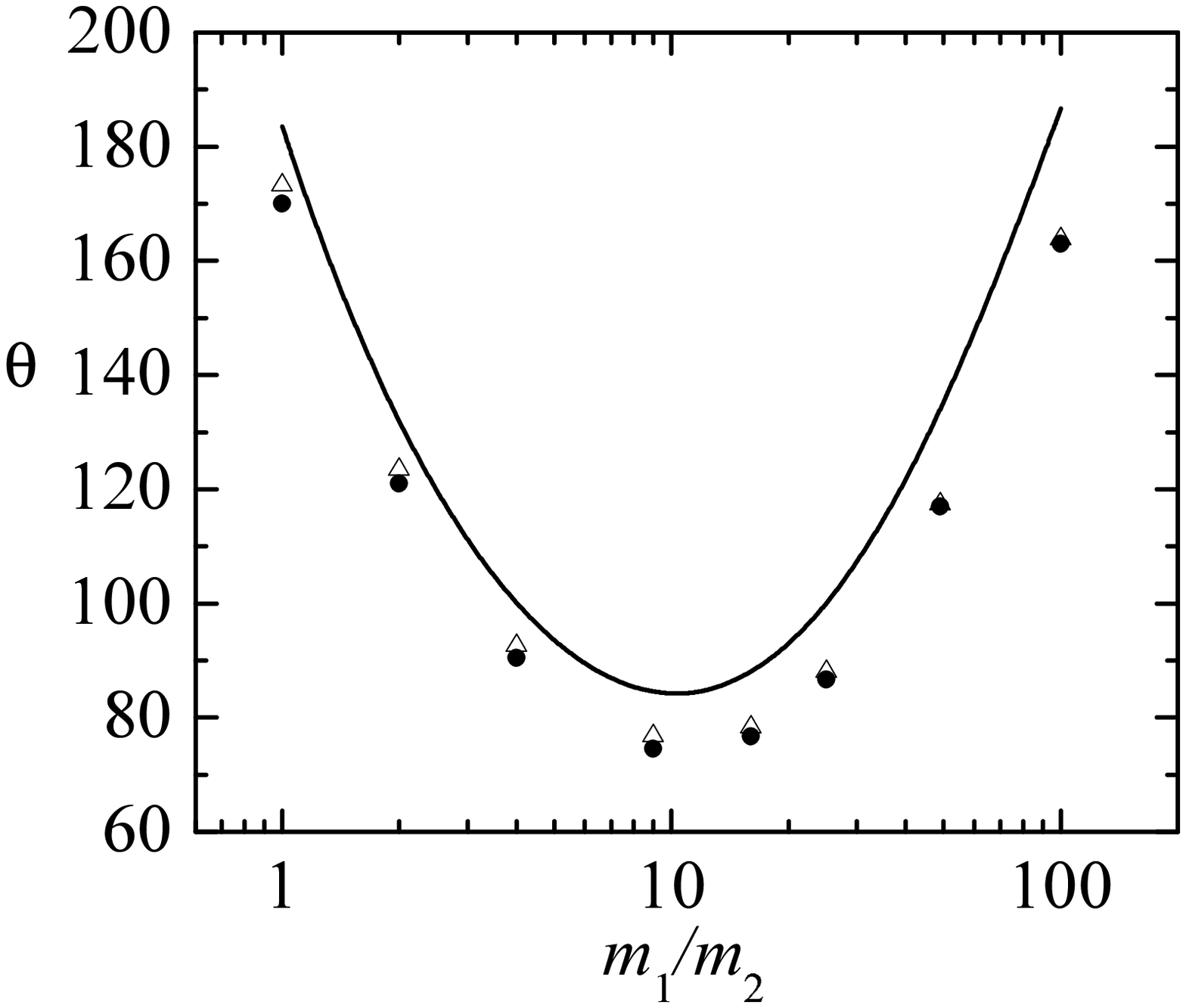}}&\resizebox{6.6cm}{!}{\includegraphics{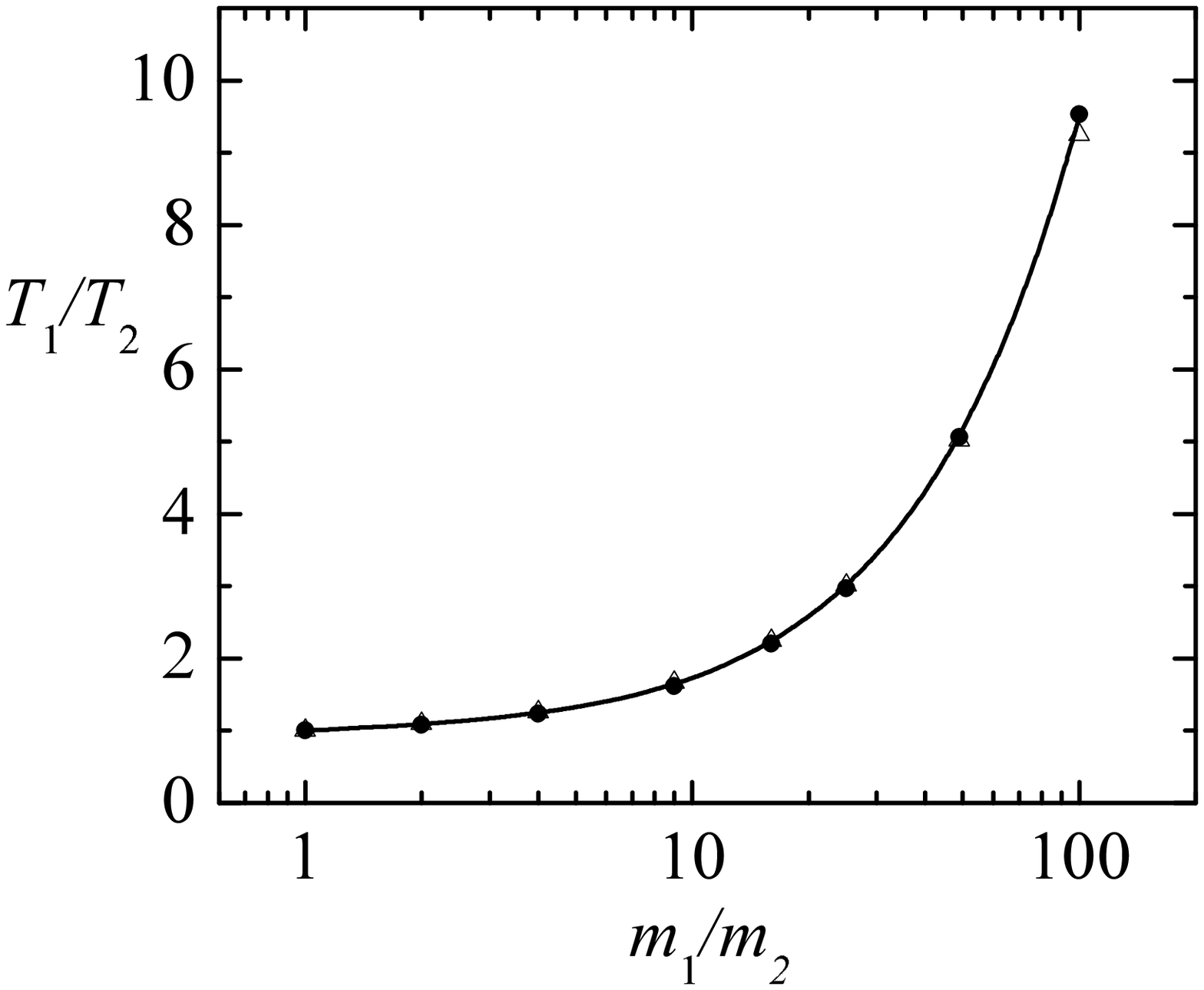}}\\
~\vspace{-0.6cm}\\
\end{tabular}
\end{center}
\caption{Reduced pressure $\Pi$, reduced shear viscosity $\mu$,
reduced temperature $\theta$ and temperature ratio $T_1/T_2$
plotted against the mass ratio $m_1/m_2$ for an equimolar mixture
($x_1=\frac{1}{2}$) of hard disks ($d=2$) with $\sigma_1/\sigma_2=1$,
$\phi=0.05$ and $\alpha=0.9$. The lines are the analytical results
obtained from the Boltzmann kinetic equation while the symbols
correspond to the ESMC results (circles) and MD simulations (triangles).
\label{fig1}}
\end{figure}

As said before, in the steady state and for given values of the parameters of the mixture,
the (dimensionless) rheological properties $\mu$, $\theta$, and $\Pi$ are functions of the coefficients
of restitution $\alpha_{ij}$ only, and their dependences on the shear-rate
are scaled out. The parameters of the mixture are the mass ratio $m_1/m_2$, the mole fraction $x_1=n_1/n$,
the ratio of diameters $\sigma_1/\sigma_2$, and the total solid volume fraction
$\phi=\phi_1+\phi_2$ where $\phi_i=n_im_i/\varrho_i$ is the species volume fraction of
the component $i$. In the following,
our aim is to examine the dependence of $\mu$, $\theta$, $\Pi$, and $T_1/T_2$ on
dissipation as well as on the parameters of the mixture,
and make a detailed comparison between theory and simulations.

\section{Results}
\label{sec3}

\subsection{2D results}
\label{sub2D}
\begin{figure}
\begin{center}
\begin{tabular}{lr}
\resizebox{6.2cm}{!}{\includegraphics{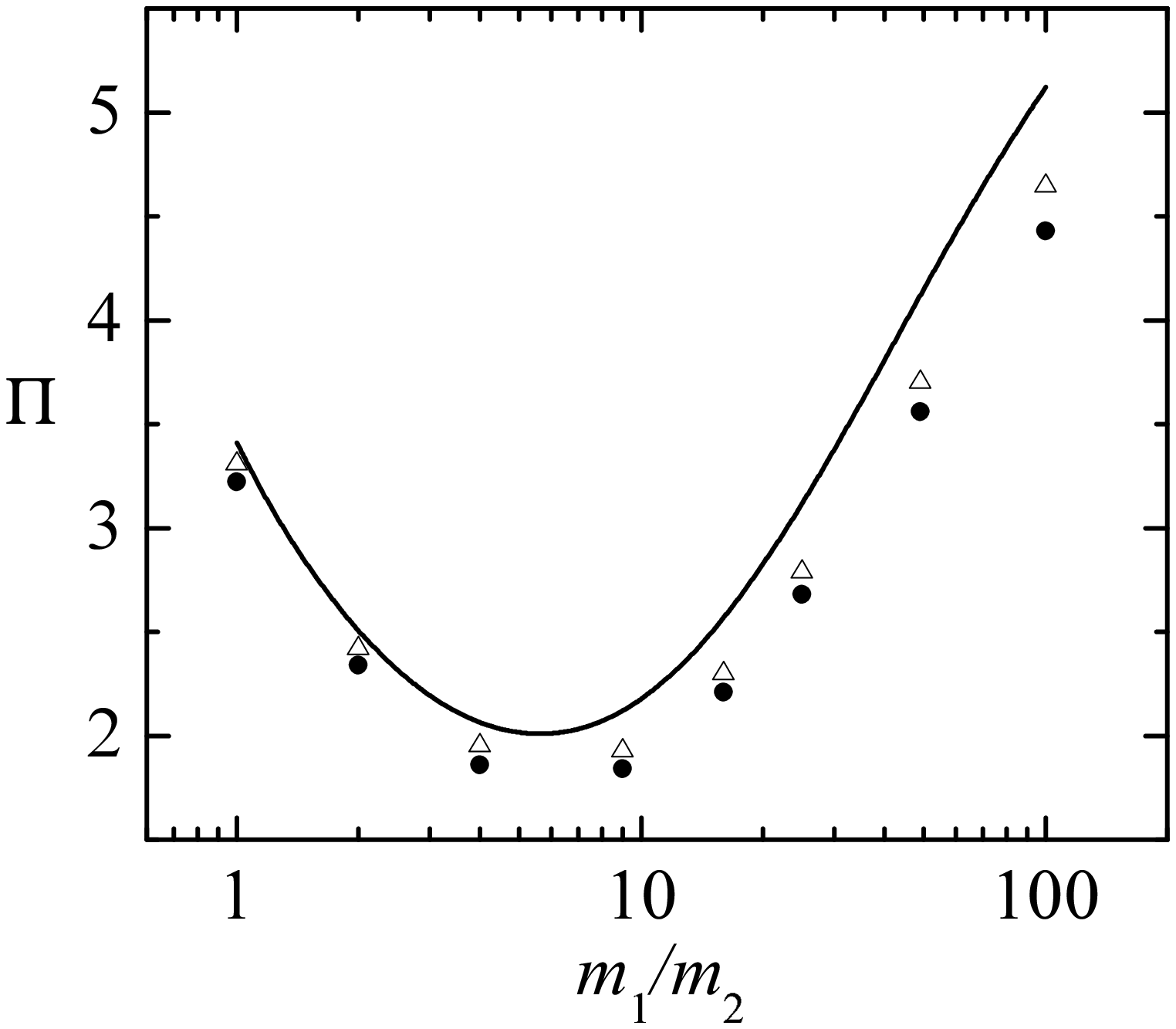}}&\resizebox{6.7cm}{!}{\includegraphics{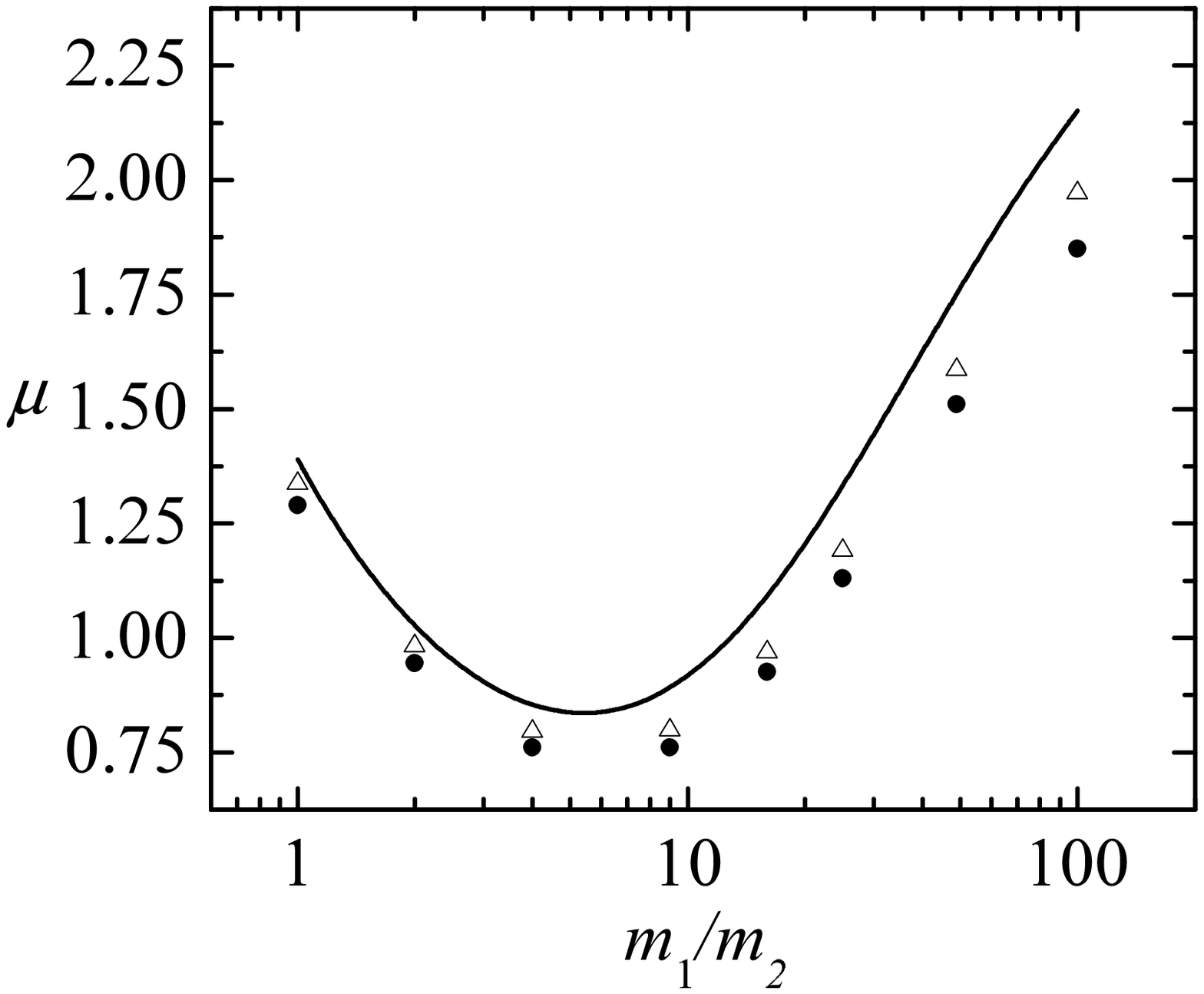}}\\
\resizebox{6.5cm}{!}{\includegraphics{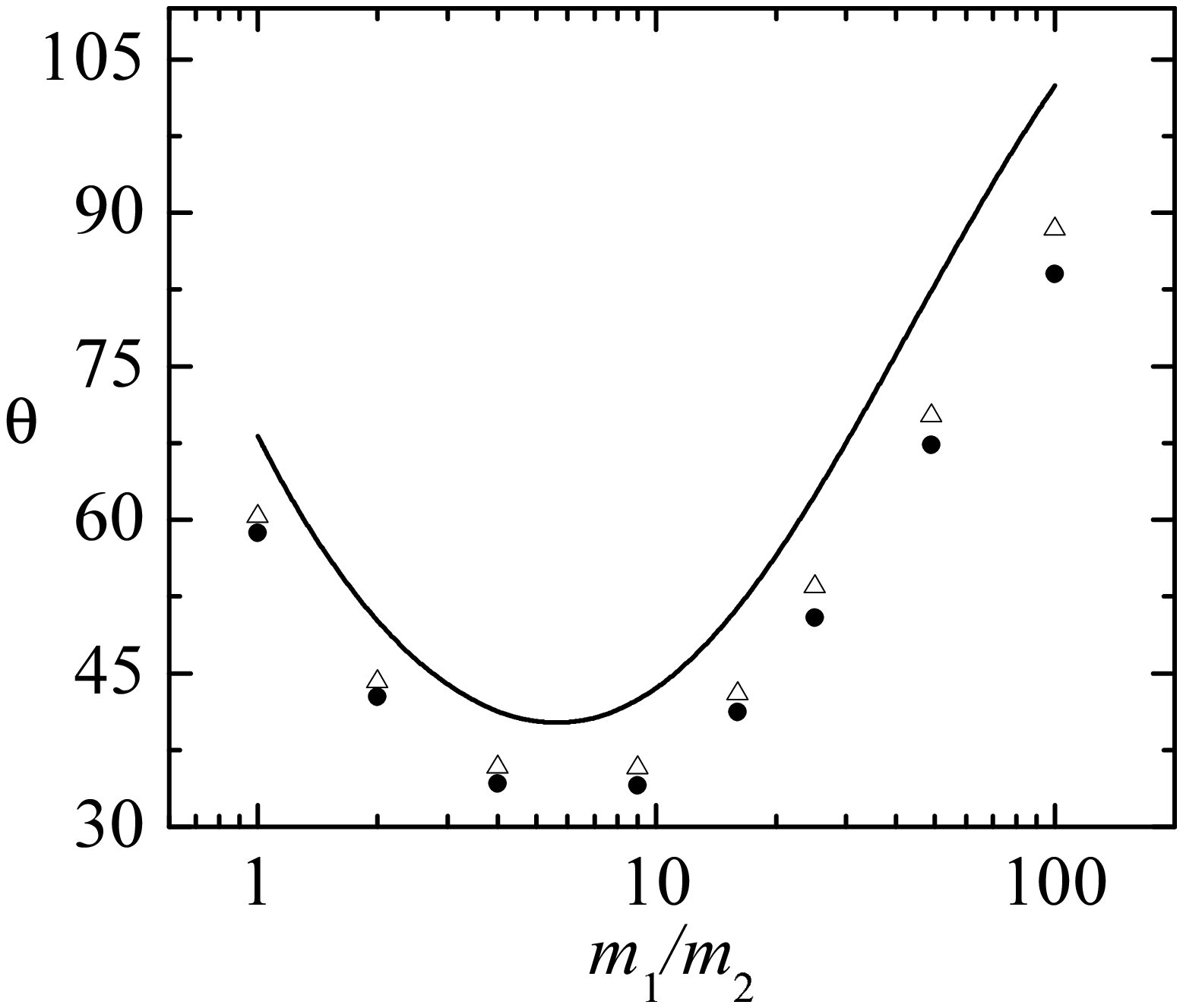}}&\resizebox{6.8cm}{!}{\includegraphics{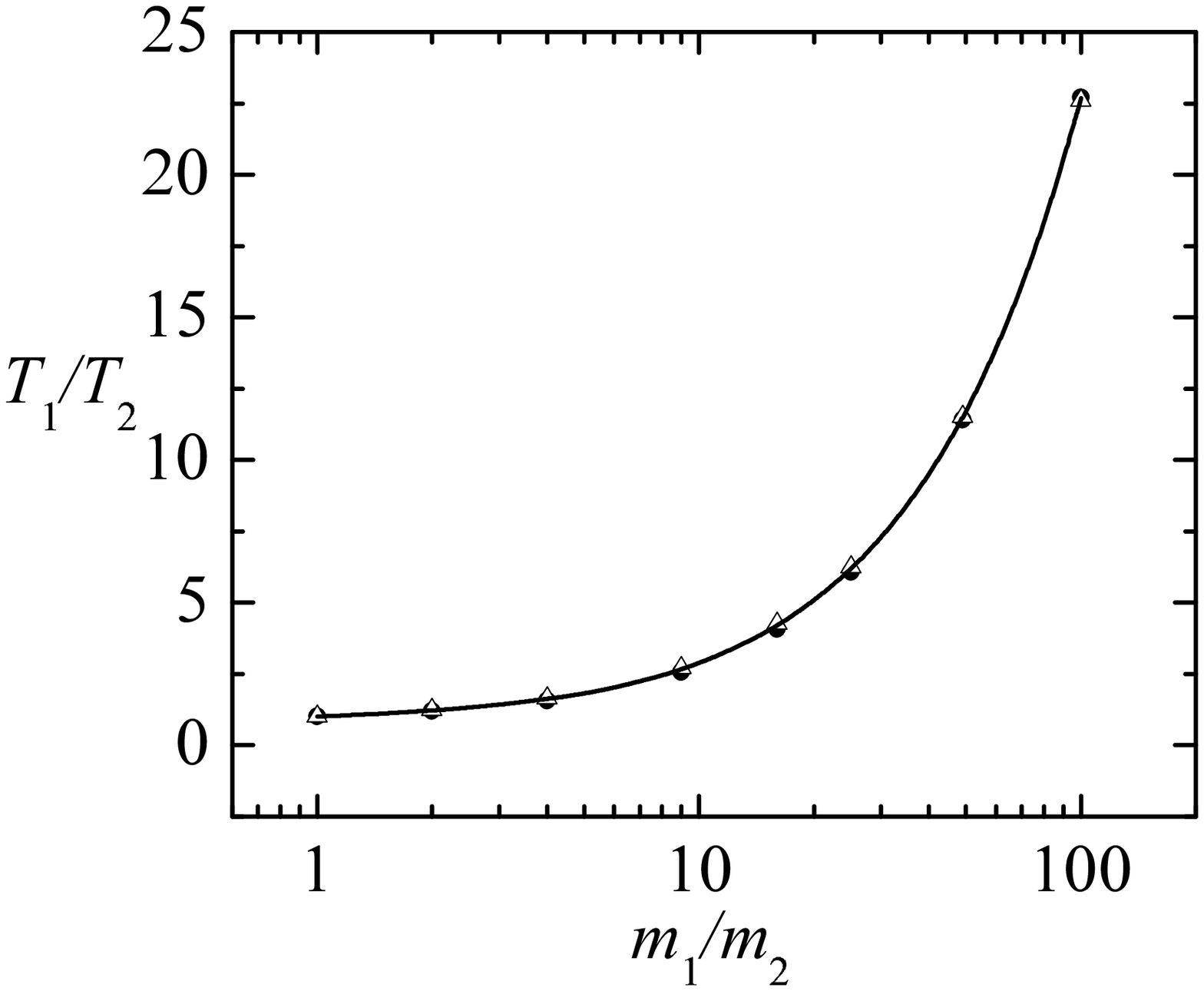}}\\
~\vspace{-0.6cm}\\
\end{tabular}
\end{center}
\caption{The same as Fig.\ \protect\ref{fig1}, only with
$\alpha=0.75$.  Note the different axis scaling as compared to
Fig.\ \protect\ref{fig1}.
\label{fig2}}
\end{figure}
The main objective of this Section is to compare the results obtained from the ESMC method
with those obtained from MD simulations performed for bidisperse mixtures of smooth hard disks ($d=2$).
For the sake of simplicity, we assume that all the coefficients of restitution are set equal
($\alpha_{ij}\equiv \alpha$) so that the parameter space of the problem is reduced to five
quantities: $\{\alpha, \phi, m_1/m_2, \sigma_1/\sigma_2, x_1\}$, where $m_1\geq m_2$ and $\sigma_1\geq \sigma_2$
is implied in the following.

For the pair correlation function values at contact, $\chi_{ij}$, the
relation \cite{JM87}
\begin{equation}
\label{5.1}
\chi_{ij}=\frac{1}{1-\phi}+\frac{9}{16}\frac{\beta}{(1-\phi)^2}\frac{\sigma_i\sigma_j}{\sigma_{ij}},
\end{equation}
is used in the Enskog and ESMC calculations,
where $\beta=\pi(n_1\sigma_1+n_2\sigma_2)/4$.
The approximation (\ref{5.1}) for the equilibrium pair correlation function is
accurate in most of the fluid region, although a more detailed comparison with computer
simulations shows that the expression becomes less accurate
with increasing density and diameter ratios \cite{LS04}.
However, given the values considered in our simulations, we expect that these approximations to
$\chi_{ij}$ turn out to be reliable with an error margin of much less than one per-cent,
so that we do not consider the higher order corrections here.
\begin{figure}
\begin{center}
\begin{tabular}{lr}
\resizebox{6.2cm}{!}{\includegraphics{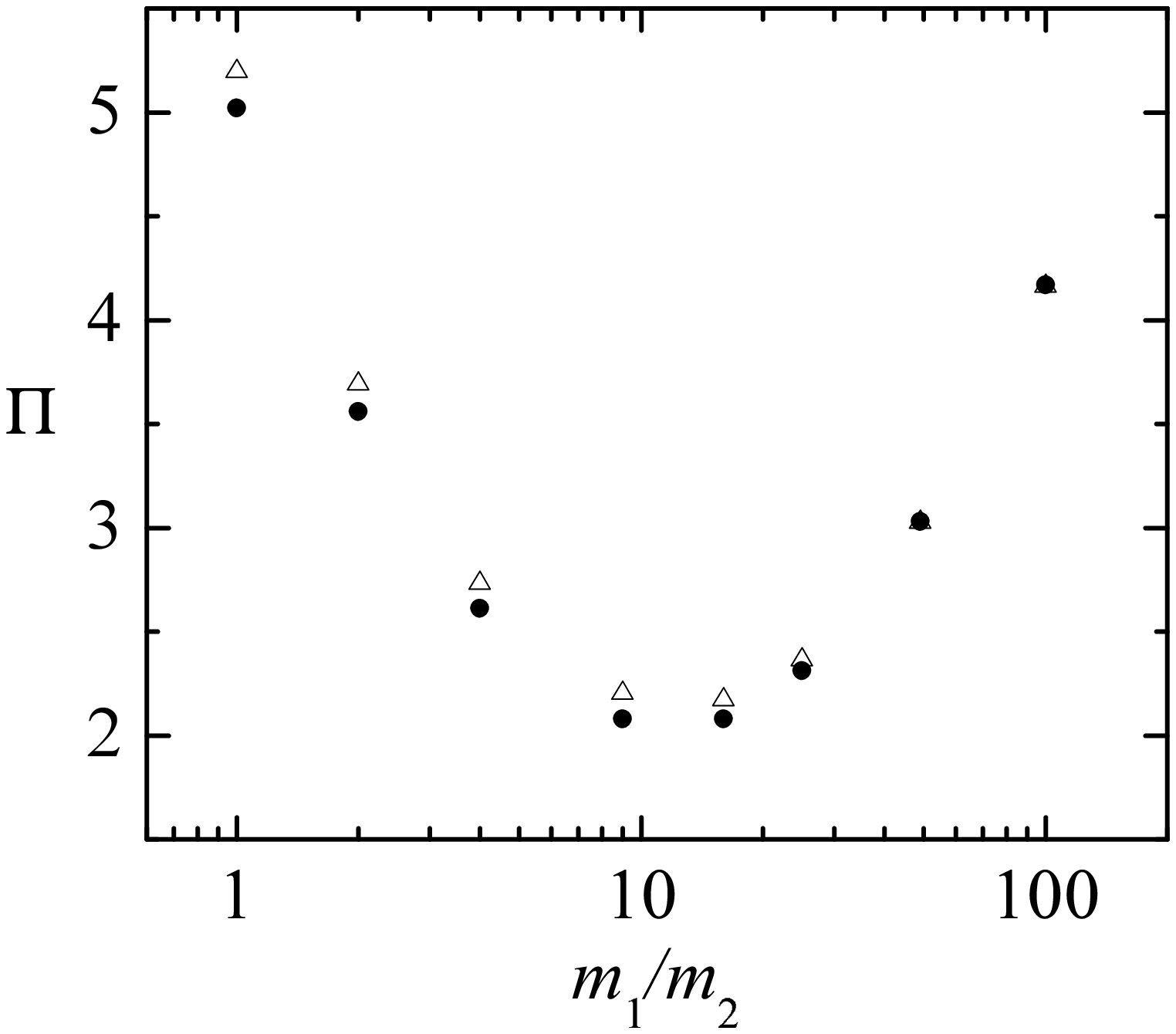}}&\resizebox{6.7cm}{!}{\includegraphics{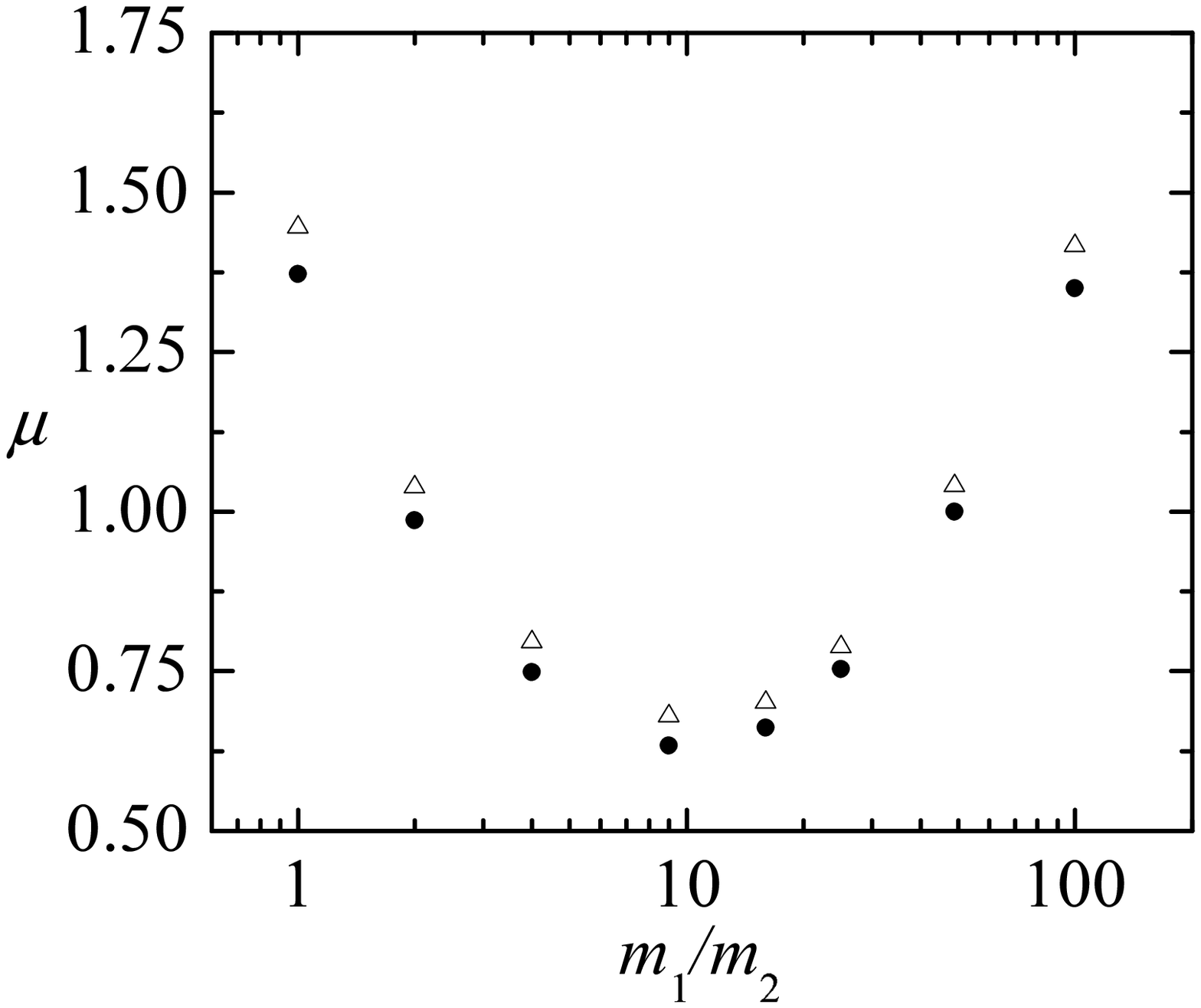}}\\
\resizebox{6.5cm}{!}{\includegraphics{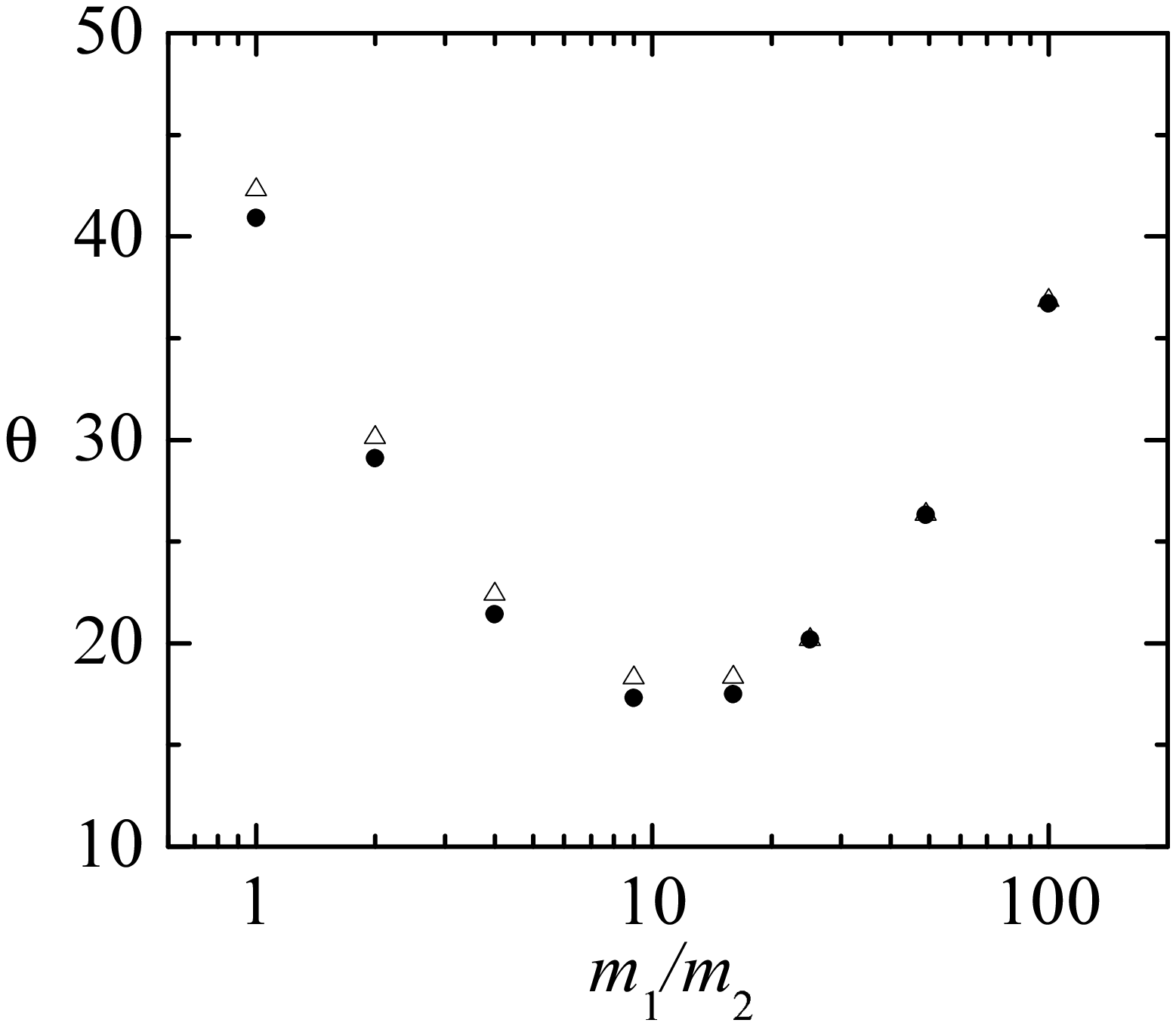}}&\resizebox{6.8cm}{!}{\includegraphics{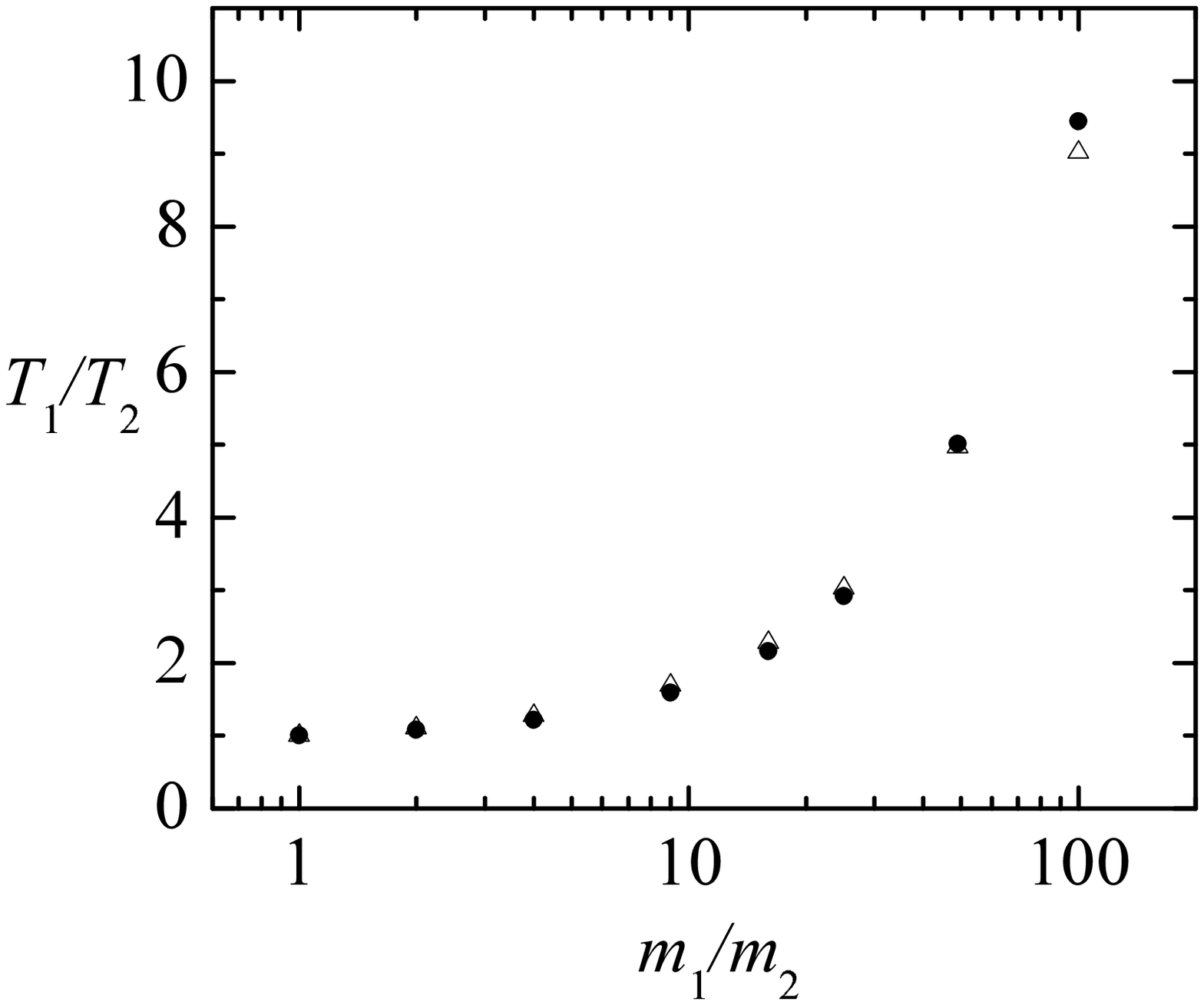}}
\end{tabular}
\end{center}
\caption{The same as Fig.\ \protect\ref{fig1}, only with $\phi=0.1$.
\label{fig3}}
\end{figure}

\begin{figure}
\begin{center}
\begin{tabular}{lr}
\resizebox{6.2cm}{!}{\includegraphics{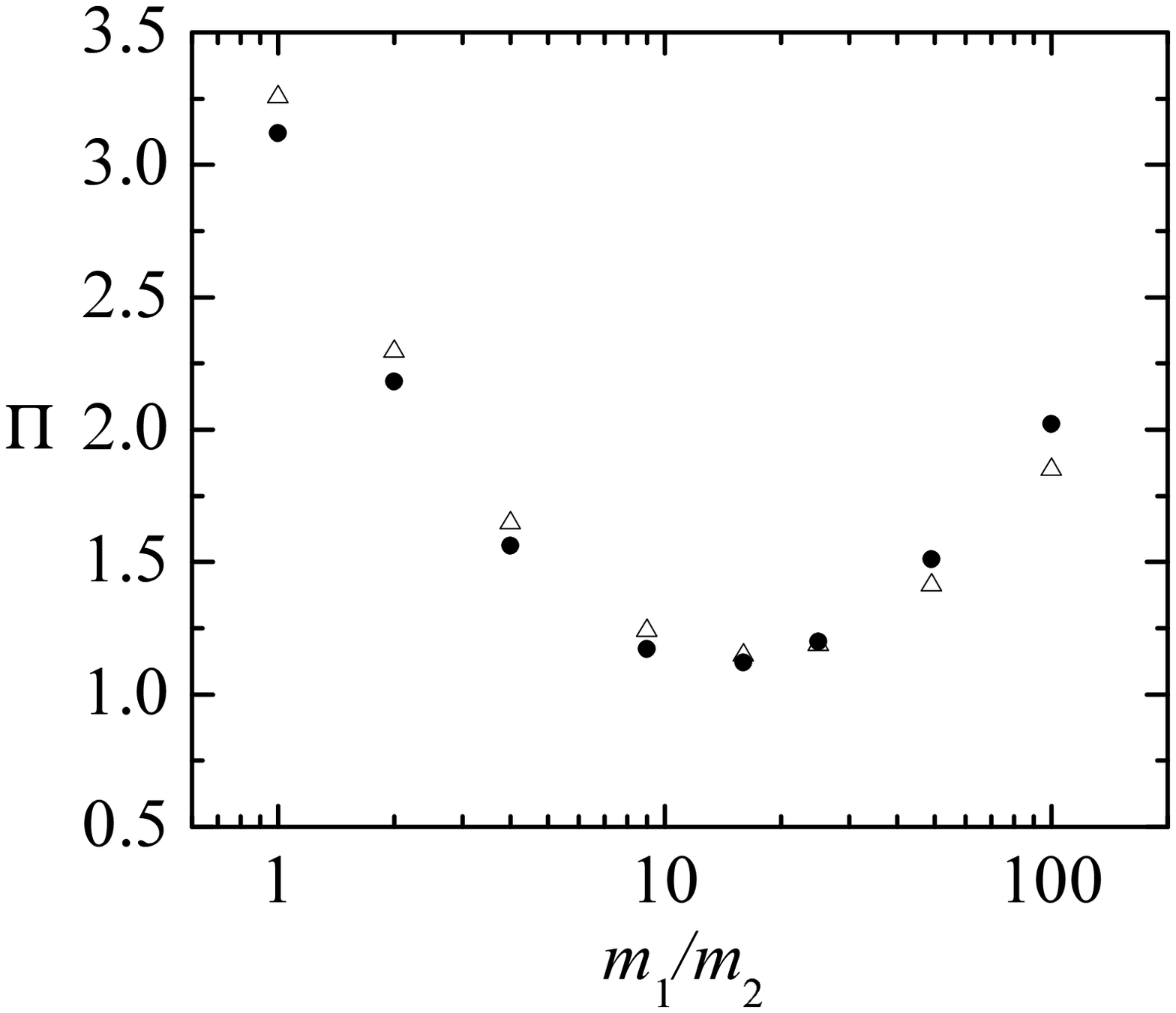}}&\resizebox{6.3cm}{!}{\includegraphics{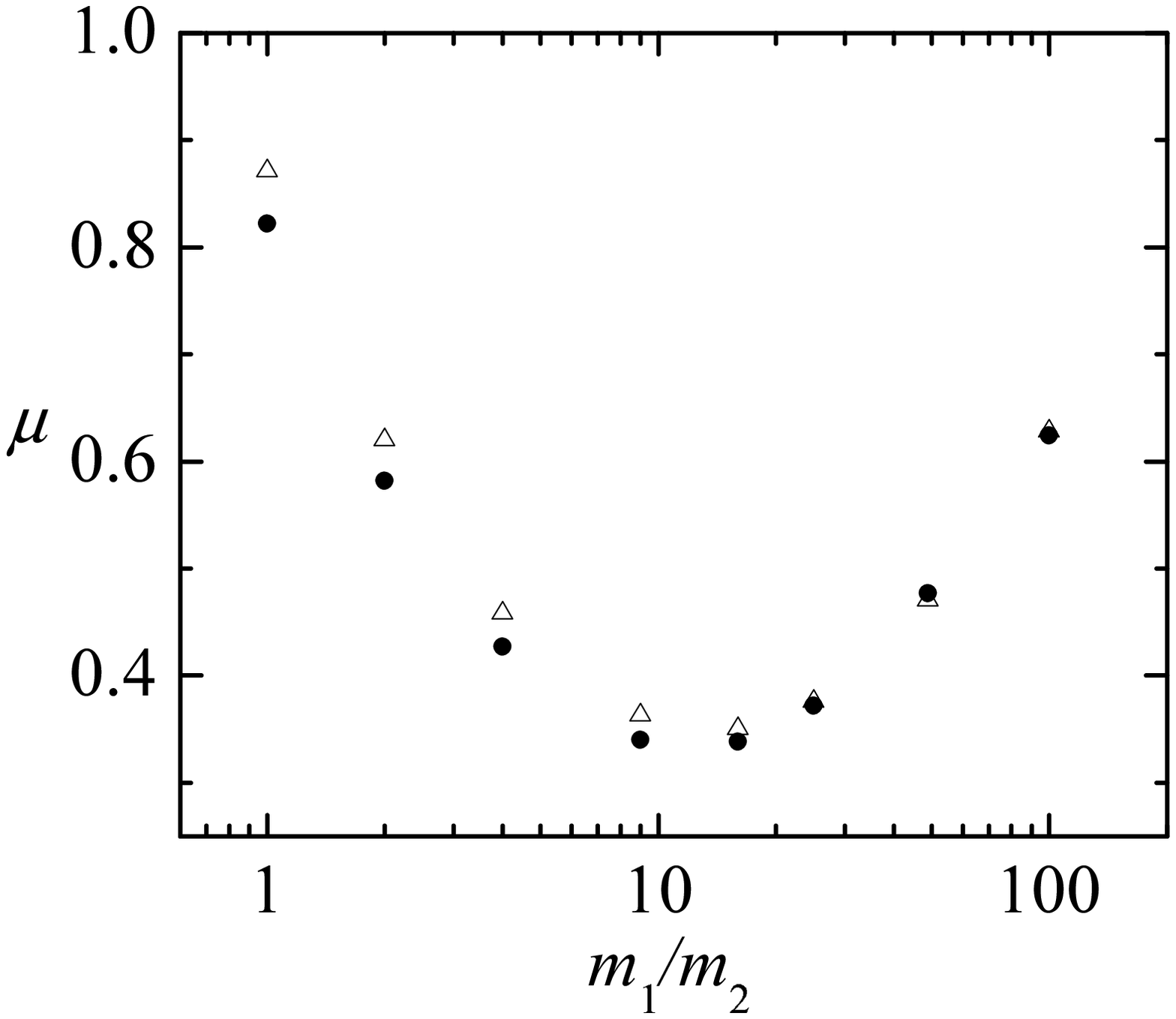}}\\
\resizebox{6.3cm}{!}{\includegraphics{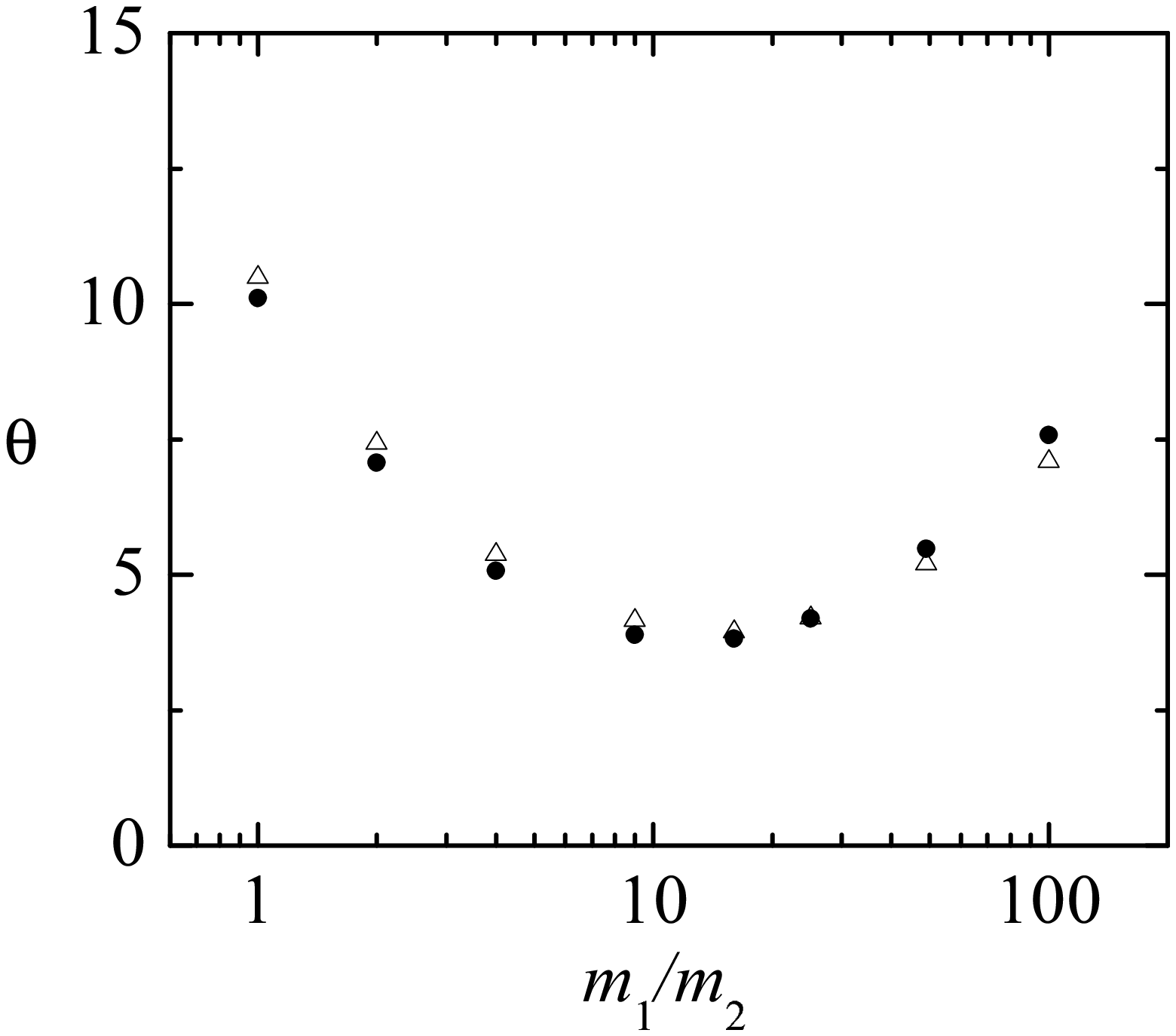}}&\resizebox{6.8cm}{!}{\includegraphics{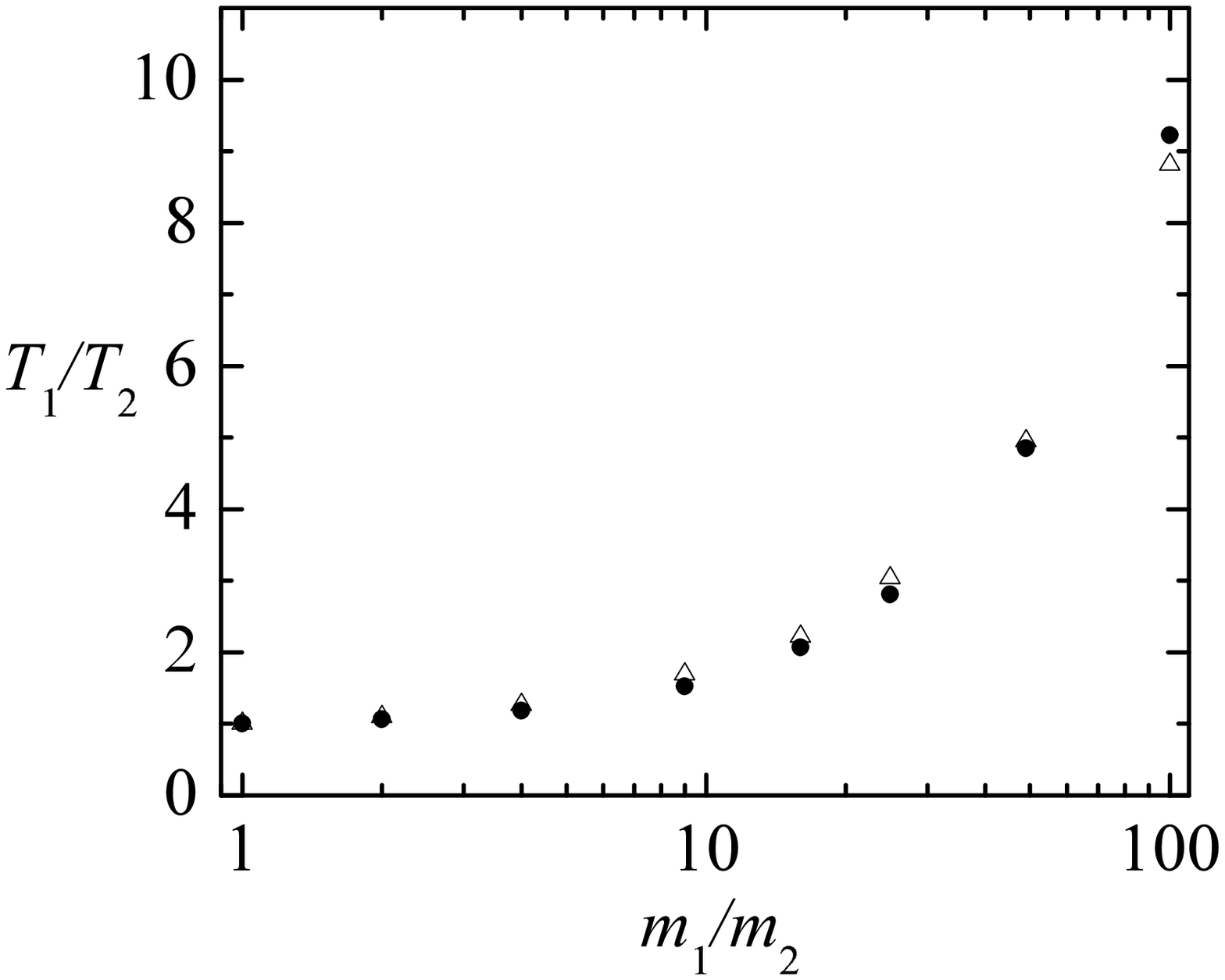}}
\end{tabular}
\end{center}
\caption{The same as Fig.\ \protect\ref{fig1}, only with $\phi=0.2$.
\label{fig4}}
\end{figure}

\begin{figure}
\begin{center}
\begin{tabular}{lr}
\resizebox{6.2cm}{!}{\includegraphics{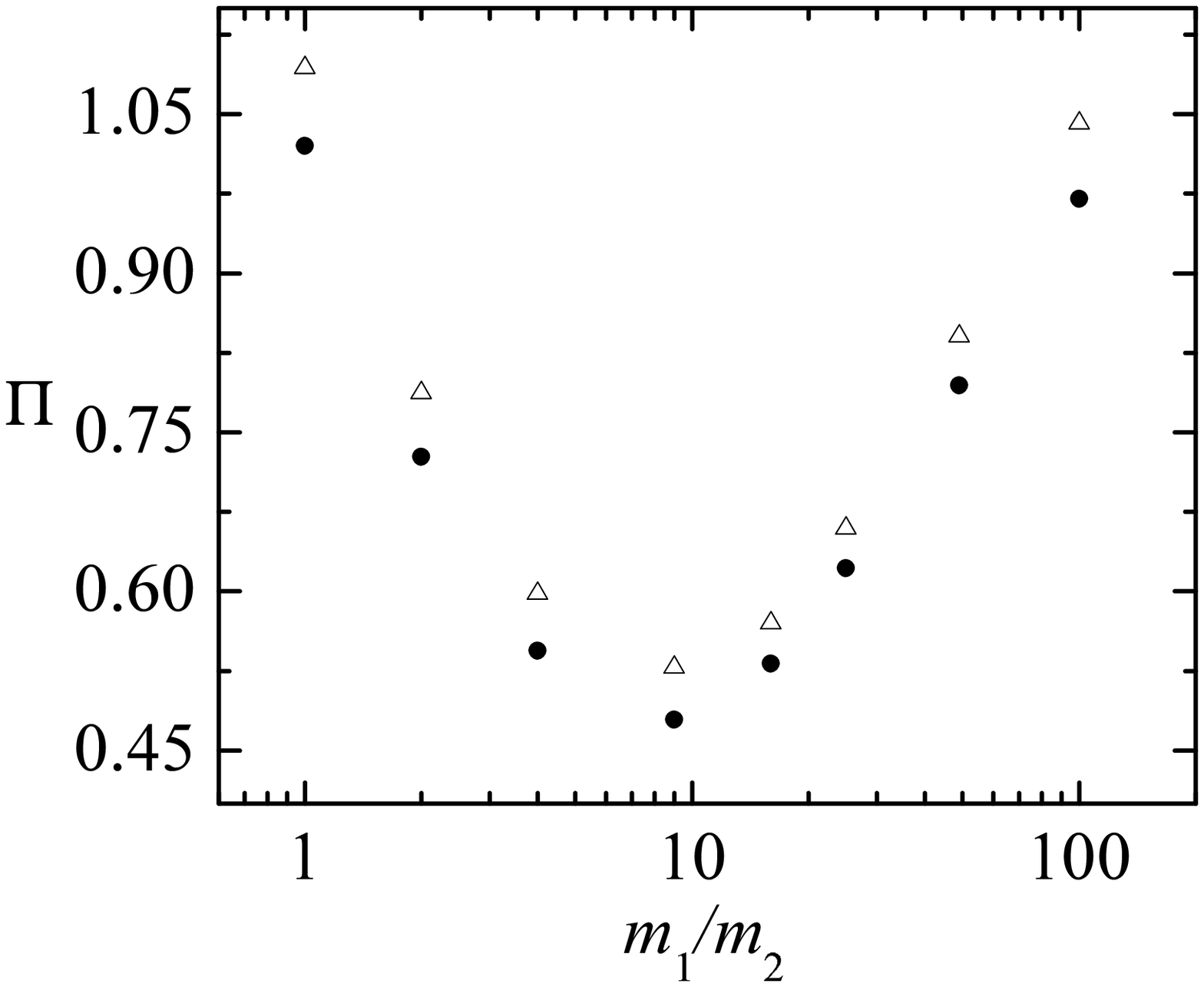}}&\resizebox{6.3cm}{!}{\includegraphics{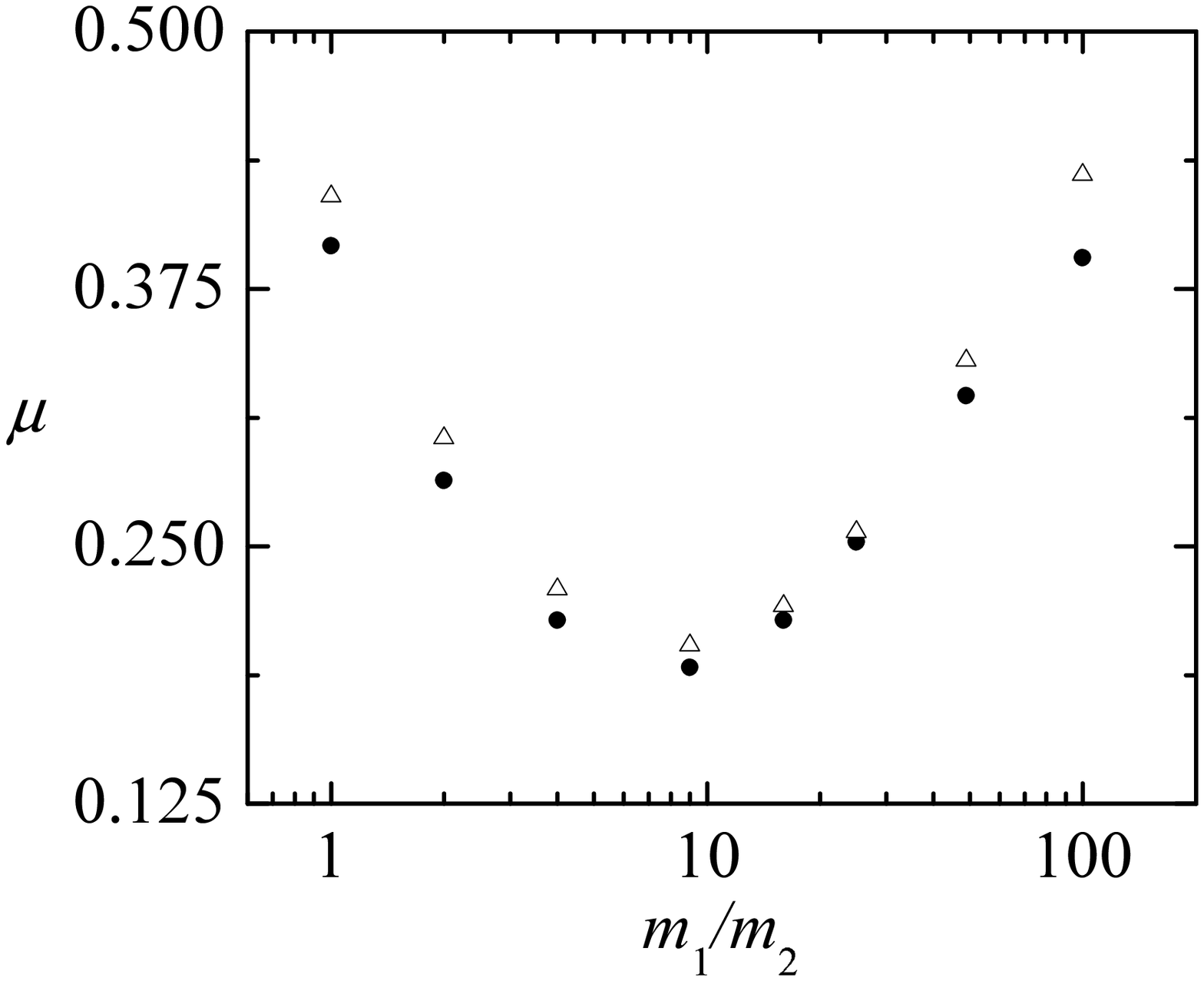}}\\
\resizebox{6.3cm}{!}{\includegraphics{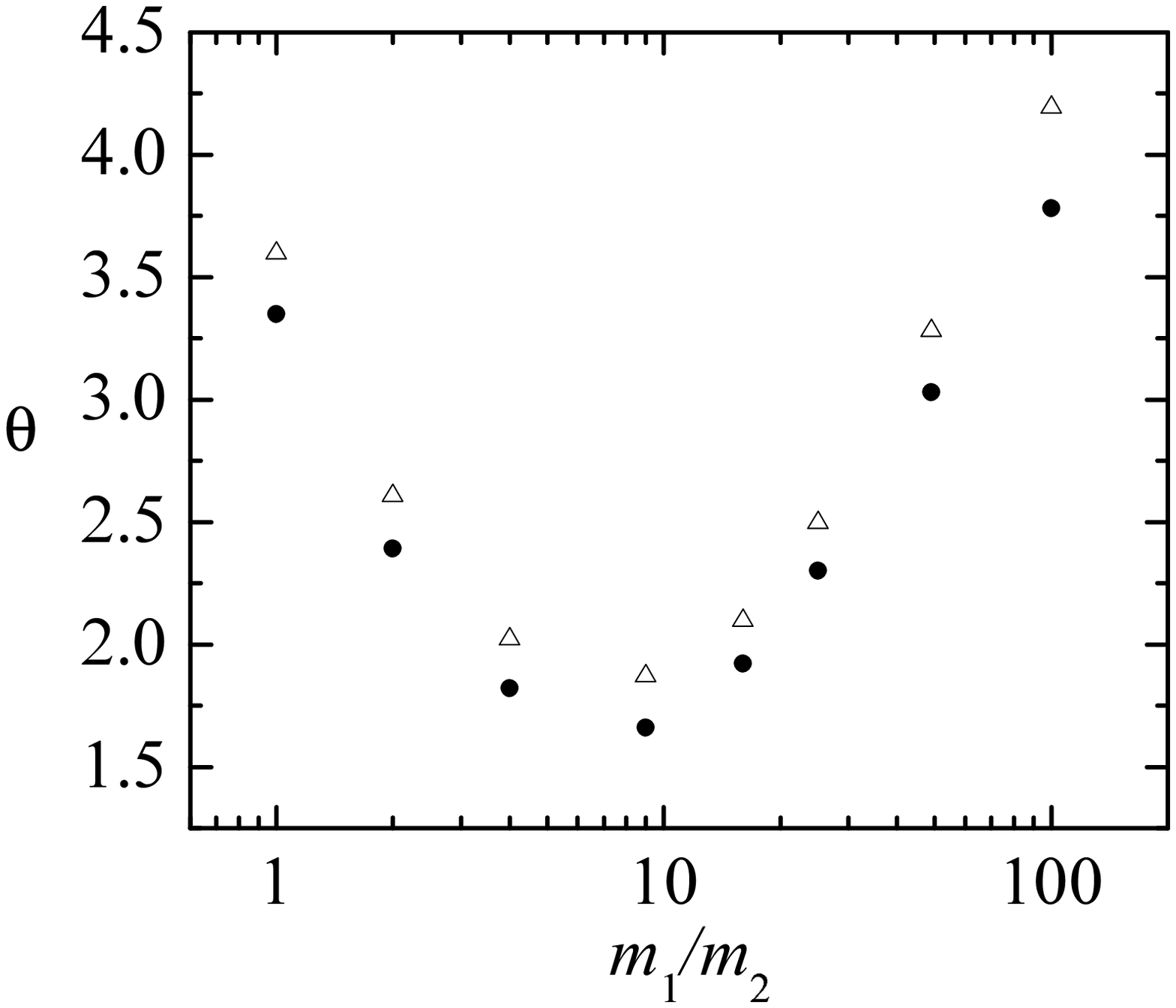}}&\resizebox{6.8cm}{!}{\includegraphics{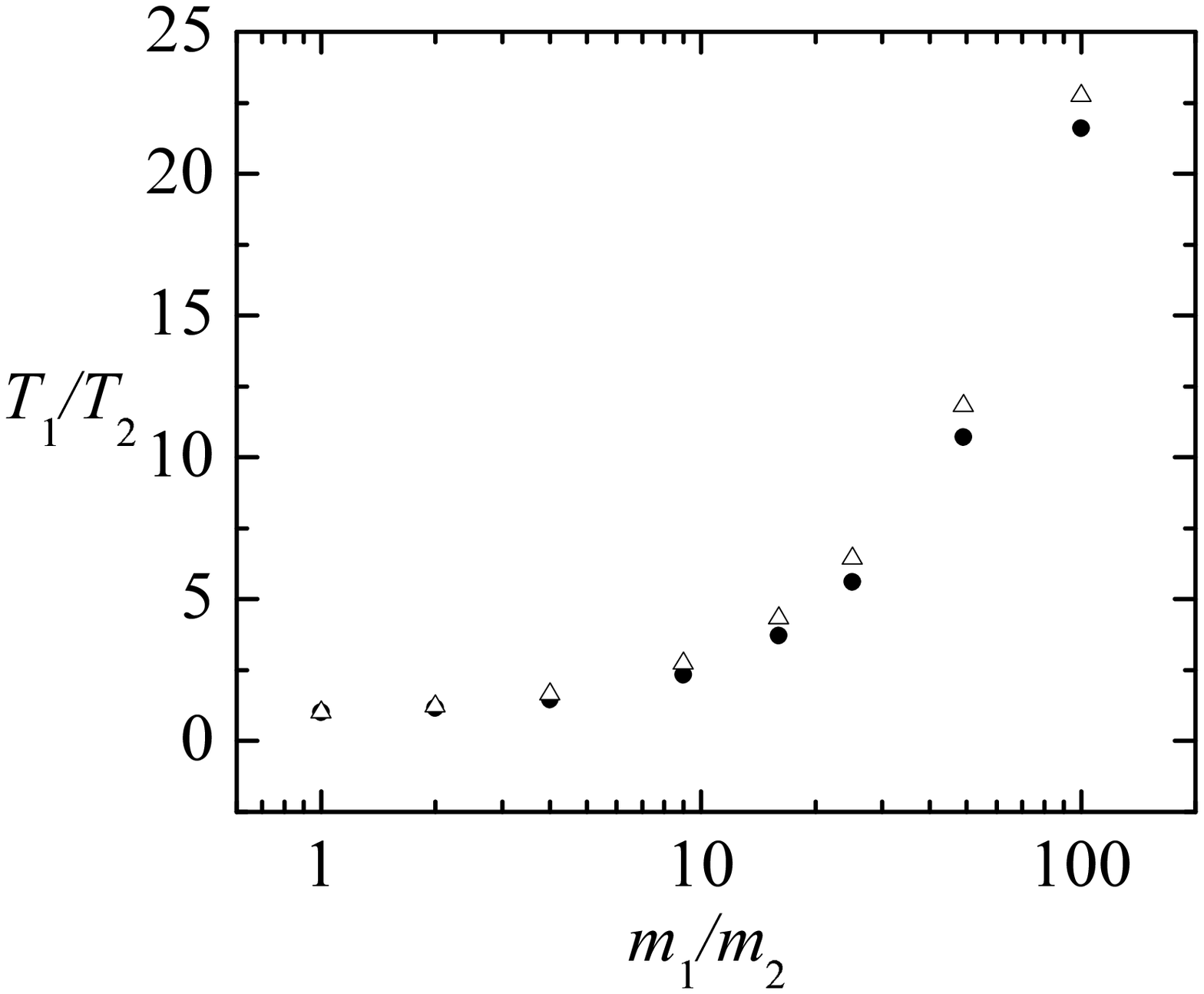}}
\end{tabular}
\end{center}
\caption{The same as Fig.\ \protect\ref{fig1}, only with
$\phi=0.2$ and $\alpha=0.75$.
\label{fig5}}
\end{figure}

Three different values of the solid volume fraction $\phi$ have been considered
here, $\phi=0.05$, $\phi=0.1$, and $\phi=0.2$.
The first value of $\phi$ represents
a dilute gas while the two latter can be considered as moderately dense fluids.
To compare with MD results, two values of the coefficient of restitution have
been considered, $\alpha=0.9$ and $\alpha=0.75$, both representing moderately
strong dissipation. We have taken an equimolar mixture ($x_1=\frac{1}{2}$)
with particles of equal size ($\sigma_1=\sigma_2$) but with quite different
masses. Specifically, we have studied the dependence of $\mu$, $\theta$, $\Pi$,
and the temperature ratio $T_1/T_2$ on the mass ratio $m_1/m_2$ for different
values of $\alpha$ and $\phi$. As said before, as the coefficient of restitution
decreases, the system goes away from equilibrium  and the energy equipartition
is not expected to hold.

In Figs.\ \ref{fig1} and \ref{fig2}, we plot  $\Pi$, $\mu$, $\theta$, and $T_1/T_2$ against
$m_1/m_2$ for low density, $\phi=0.05$, and the coefficients of restitution
$\alpha=0.9$ and $\alpha=0.75$, respectively.
The symbols refer to MD simulations (triangles) and
ESMC results (circles) while the solid lines correspond to the results
obtained from the Boltzmann equation \cite{GM03},
which is strictly valid in the limit $\phi\to 0$.
In Fig.\ \protect\ref{fig1}, the agreement between the
ESMC and MD results is excellent, even for very disparate masses.
The Boltzmann results compare also quite well with
simulations, especially in the case of the temperature ratio.
For the steady temperature,
the Boltzmann predictions slightly overestimate the MD simulation results.
These discrepancies can be primarily attributed to density effects, since
previous
comparisons between theory and DSMC results show better agreement for
$\phi=0$ than the one reported here  \cite{GM03bis}.
Comparing the two figures, one can conclude that the agreement between the
Boltzmann kinetic theory and the simulation data becomes worse with increasing
dissipation.
The only exception is the temperature ratio which is essentially obtained from
the diagonal
elements of the kinetic contribution to the pressure tensor.
We also observe good agreement between Enskog and MD results, except for
the largest values of the mass ratio.

\begin{figure}
\begin{center}
\begin{tabular}{lr}
\resizebox{6.2cm}{!}{\includegraphics{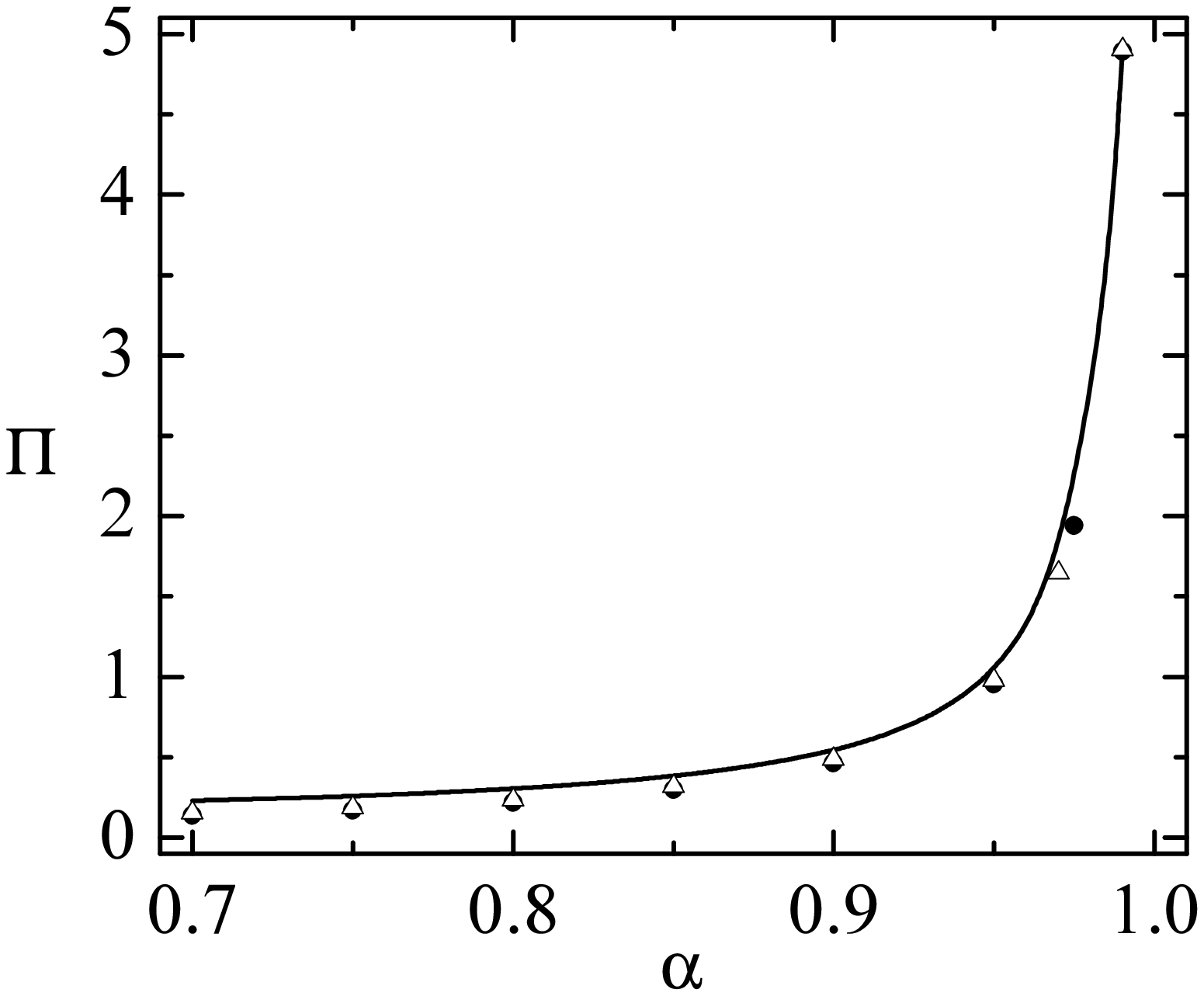}}&\resizebox{6.5cm}{!}{\includegraphics{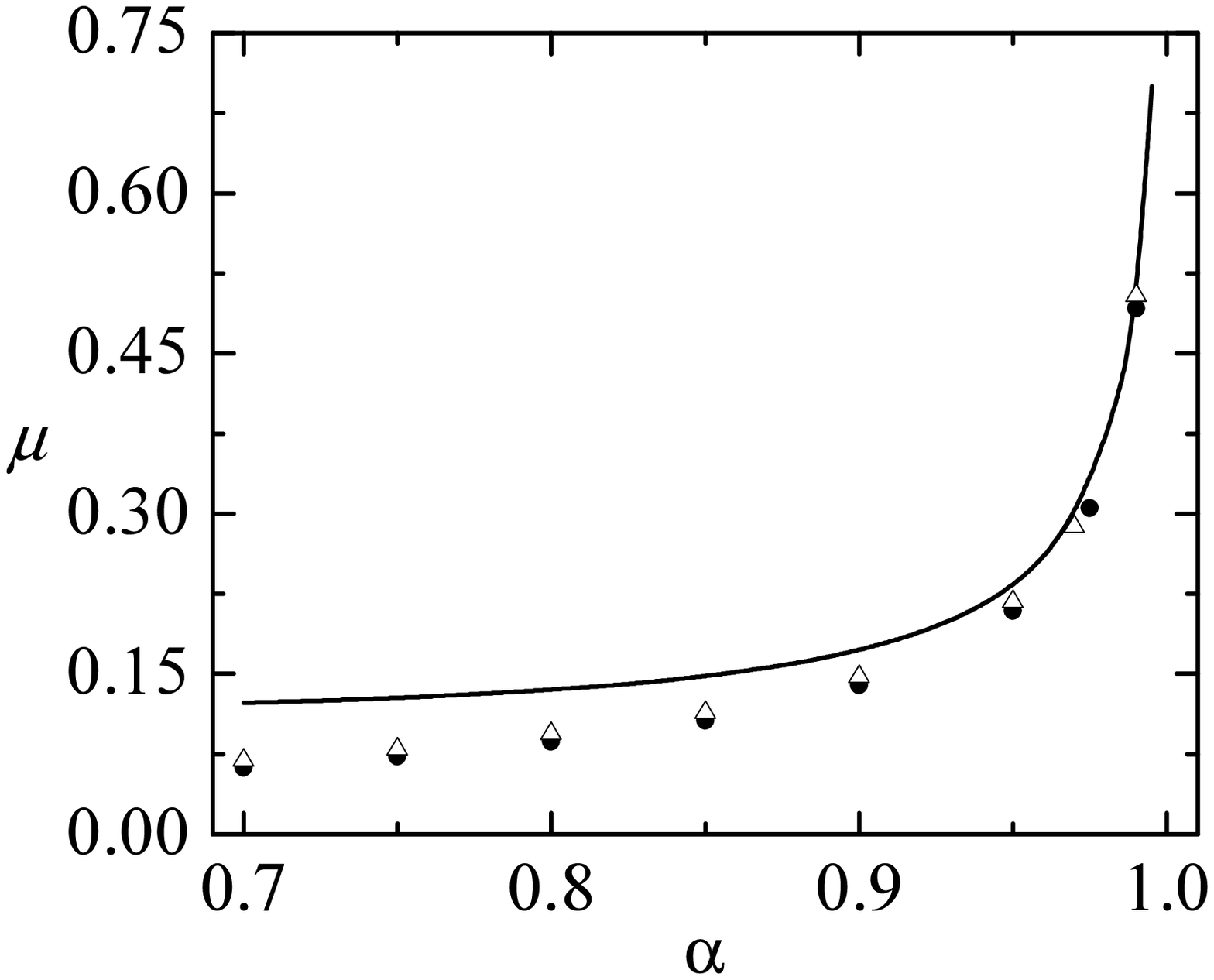}}\\
\resizebox{6.3cm}{!}{\includegraphics{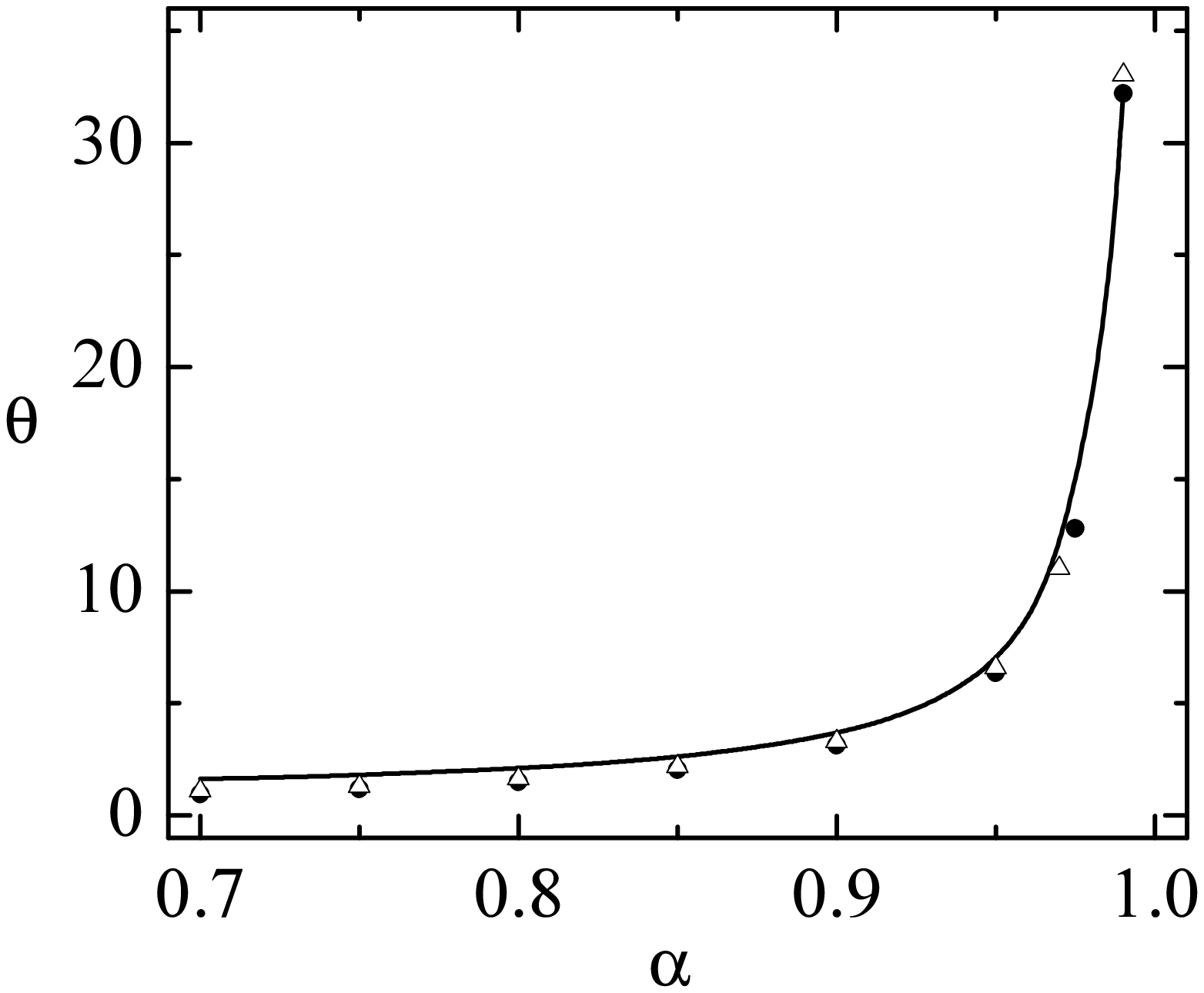}}&\resizebox{6.9cm}{!}{\includegraphics{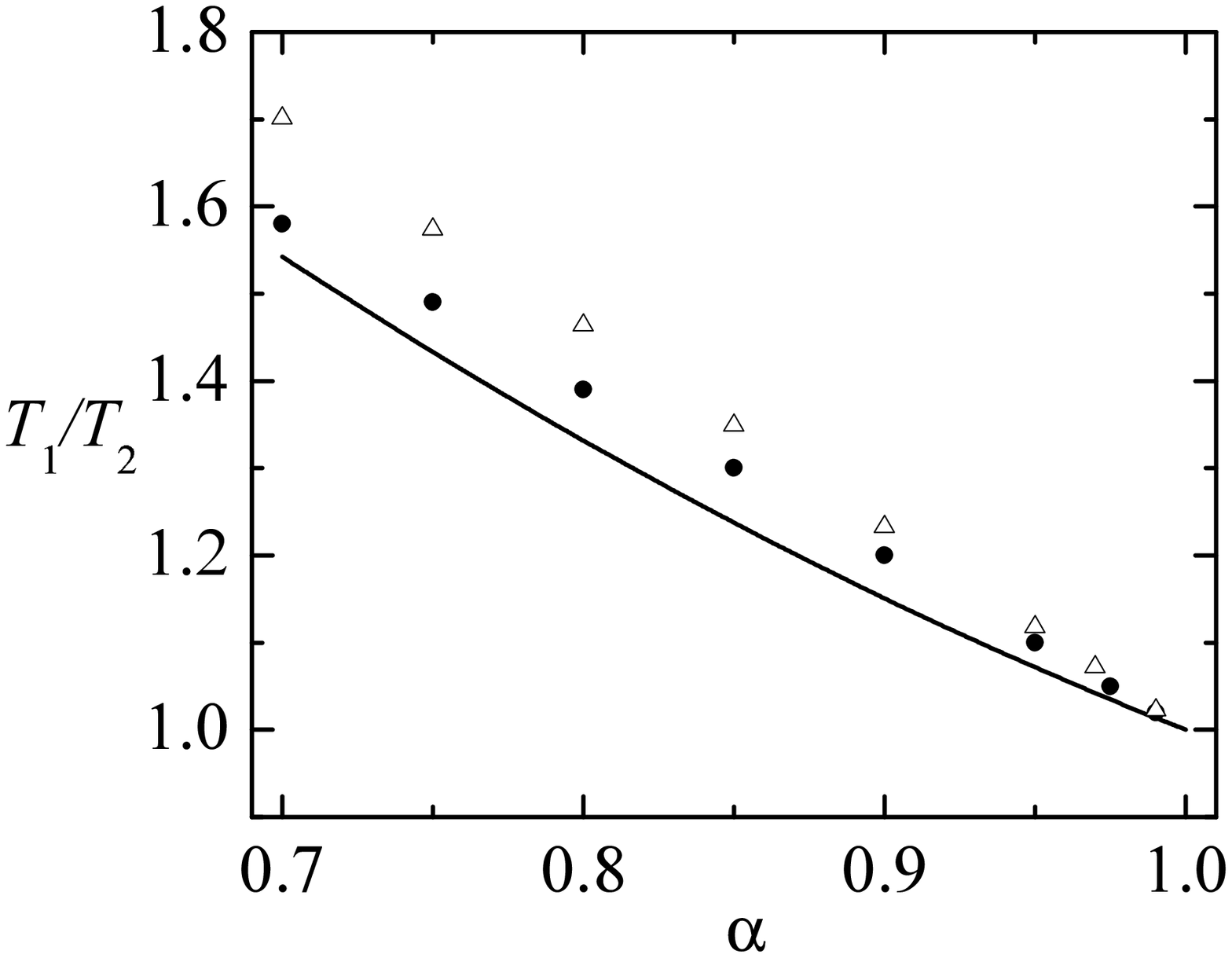}}
\end{tabular}
\end{center}
\caption{Plot of the reduced pressure $\Pi$, the reduced shear viscosity $\mu$, the reduced temperature $\theta$ and the temperature ratio $T_1/T_2$ versus the coefficient of restitution $\alpha$  for an equimolar mixture ($x_1=\frac{1}{2}$) of hard spheres ($d=3$) with $\sigma_1/\sigma_2=1$, $m_1/m_2=4$, and $\phi=0.1$. The lines are the analytical results obtained from the Chapman-Enskog solution up to the Navier-Stokes order while the symbols (circles) correspond to the ESMC results.
\label{fig6}}
\end{figure}

Figures \ref{fig3} and \ref{fig4} depict results for the higher densities
$\phi=0.1$ and $\phi=0.2$, with a coefficient of restitution $\alpha=0.9$.
The Enskog (ESMC) results agree well with the MD simulations, only
does the discrepancy increase in general with
decreasing mass ratio for the pressure $\Pi$ and the temperature
$\theta$, while the opposite occurs for the temperature ratio. We also
observe that in general the Enskog kinetic theory
underestimates the MD results.
Thus, for instance in the case $m_1/m_2=4$ and for $\phi=0.2$,
the discrepancies for $\Pi$, $\mu$, $T_1/T_2$, and $\theta$ are
about $5\%$, $7\%$, $7\%$, and $5\%$, respectively.
In addition, both Enskog and MD results clearly predict a
{\em non-monotonic} behavior of the pressure and the
viscosity on the mass ratio. {This conclusion contrasts with
the predictions of kinetic theories \cite{WA99} based on the
equipartition assumption which suggest a monotonic dependence
of $\Pi$ and $\mu$ on $m_1/m_2$ \cite{AL02,GM03bis}}.
With respect to the behavior of the temperature ratio,
we observe that $T_1/T_2$ exhibits a strong dependence on the mass ratio.
The mass disparity enhances the magnitude of the
non-equipartition of energy. Differences between the Enskog theory and
MD simulations become more significant as the dissipation increases,
as shown in Fig.\ \ref{fig5} for $\phi=0.2$ and $\alpha=0.75$.
Although the Enskog theory captures well the trends observed in
MD simulations, the disagreement between both approaches is more
important than the one observed in previous cases.
More specifically, in the case $m_1/m_2=4$, the discrepancies
for $\Pi$, $\mu$, $T_1/T_2$, and $\theta$ are now about
$9\%$, $6\%$, $11\%$, and $10\%$, respectively. However, given that
these deviations are not quite large, one can still consider the
Enskog equation  as a good approximation for describing the
rheological properties of a granular mixture, even for moderately
high density and strong dissipation.

\subsection{3D results}

In the case of hard spheres ($d=3$), we take for the pair correlation function
$\chi_{ij}$ the following approximation \cite{CS}
\begin{equation}
\label{5.2}
\chi_{ij}=\frac{1}{1-\phi}+\frac{3}{2}\frac{\beta}{(1-\phi)^2}\frac{\sigma_i\sigma_j}{\sigma_{ij}}
+\frac{1}{2}\frac{\beta^2}{(1-\phi)^3}\left(\frac{\sigma_i\sigma_j}{\sigma_{ij}}\right)^2,
\end{equation}
where now $\beta=\pi(n_1\sigma_1^2+n_2\sigma_2^2)/6$.
We compare the ESMC and ED results
obtained for a mixture under USF with a kinetic theory
recently proposed \cite{GM03}. In contrast to previous theories
\cite{WA99,AL02b}, the above theory takes into account the effect
of non-equipartition of granular energy on the transport
coefficients. This theory is based on the Chapman-Enskog solution
\cite{CC70} to the Enskog equation in the first order of
the shear rate (Navier-Stokes approximation).
However, given that USF is inherently non-Newtonian, the full
nonlinear dependence of the viscosity on the shear rate is required.
This implies that there is no possibility {\em a priori} of using
the Navier-Stokes equations to describe the (shear-rate dependent)
rheological properties of USF, especially as the dissipation
increases \cite{SGD04}. In Fig.\ \ref{fig6} we plot the reduced
quantities $\Pi$, $\mu$, $\theta$, and $T_1/T_2$ as a
function of the coefficient of restitution $\alpha$ for
$\sigma_1/\sigma_2=1$, $x_1=\frac{1}{2}$, $m_1/m_2=4$,
and $\phi=0.1$ as given by Monte Carlo simulations (circles),
MD simulations (triangles)
and the Chapman-Enskog solution (solid lines). Given that in reduced units,
the shear rate and dissipation are not independent parameters,
one would expect that the differences between the Navier-Stokes
predictions and the simulation results increase as the coefficient
of restitution decreases. Figure \ref{fig6} confirms the above
expectations since, for instance, the discrepancies between
theory and MD simulations for $\Pi$, $\mu$, $\theta$,
and $T_1/T_2$ are about 30\%, 43\%, 32\%, and 10\%, respectively,
at $\alpha=0.7$. This shows again
that the hydrodynamic description of the USF
state is non-Newtonian.  Nevertheless,
the main trends observed for the rheological properties
are qualitatively well captured by the Chapman-Enskog solution.
Concerning the comparison between
the ESMC and MD results we observe good agreement, except
for strong dissipation where the discrepancies are significant
(say, for instance, about 10\% for $\alpha=0.7$).

\section{Discussion}
\label{sec4}

In an effort to validate the Enskog kinetic theory,
the Enskog equation for a binary mixture
of smooth {\em inelastic} hard disks/spheres
under uniform shear flow (USF), in steady state,
has been numerically solved by means of
the ESMC method \cite{MS96}.
This method can be considered as the extension of the
well-known DSMC method \cite{B94} of the Boltzmann equation for finite
densities. Results for the (reduced) rheological properties and the
temperature ratio have been reported in a wide parameter space
(mass ratio, density, and coefficients of restitution).
Variations of the diameter ratio and the composition were
studied elsewhere \cite{AL03}. In addition,
the ESMC results have been compared to event-driven molecular dynamics (MD)
for hard disks/spheres as well as to solutions of the Boltzmann equation
in the low density limit \cite{MG02} and of the
Enskog equation for small shear rates \cite{GM03}.

We have considered mixtures constituted by particles of equal size and
have basically focused on studying the effect of mass disparity on the
rheological behavior of sheared 2D granular mixtures and the effect of
dissipation on the rheological behavior of sheared 3D granular mixtures.
For the 2D simulations, three different values for the solid volume fraction
($\phi$=0.05, 0.1, and 0.2) and two values of the coefficient of restitution
($\alpha$=0.9 and 0.75) were considered.  In the case of $\alpha$=0.9,
the comparison shows that there are no significant qualitative differences
between the results obtained from MD simulations and those that follow from
the ESMC method.
As the dissipation increases ($\alpha=0.75$), although the Enskog predictions
compare qualitatively well with MD data, the discrepancies between both
approaches become more prominent.
These discrepancies tend to increase also with increasing density.
On the other hand, the good agreement found here includes densities well outside
of the Boltzmann limit and values of dissipation that clearly lie outside
of what can be considered the quasielastic limit. This test may be taken
again as a further testimony to the usefulness of the Enskog equation
for fluids with elastic and inelastic collisions, including mixtures.

Like the Boltzmann equation, the Enskog kinetic theory \cite{MG02,GM03}
assumes the molecular chaos hypothesis and thus
the two particle velocity correlations for a pair of
particles at contact are neglected. However, as mentioned
in the Introduction, recent MD simulations  performed  in the
homogeneous case for driven \cite{SM01,SPM01,PTNE01} and for
undriven \cite{BM04} granular gases have observed velocity correlations
for pairs of particles that are about to collide. These short-range
velocity correlations were relevant for high densities and finite
dissipation. Consequently, some disagreement between
the Enskog theory and simulations
with increasing dissipations is expected. However, the magnitude of
disagreement with increasing dissipation is not much, and hence
the validity of the Enskog kinetic theory appears to span a much wider range
than  the limit of nearly elastic collisions.
On the other hand, the failure of the Enskog kinetic theory at high densities
is expected just as for fluids with elastic
collisions \cite{VE90}. This is possibly due to multiparticle collisions,
including recollision events (ring collisions).

There is some indication that the effects of correlated collisions are
enhanced at strong
dissipation since the colliding particles tend to become more focused
\cite{SM01,SPM01,PTNE01}.
Furthermore, our results also indicate that the range of densities for
which the Enskog equation
holds decreases with increasing dissipation. This conclusion agrees
with previous findings
made for mixtures in the homogeneous cooling state \cite{DHGD02} and
for the case of the
self-diffusion coefficient  \cite{LBD02}. The specific mechanism
responsible for these
discrepancies at higher densities and its quantitative prediction
remains an open problem.

As said in the Introduction, one possible source of
discrepancies observed between Enskog results and MD
simulations is due to
the fact that while theory and ESMC results are restricted to
a homogeneous state (in the Lagrangian frame), some degree of
inhomogeneity (especially for density) evolves
in MD simulations in the steady state. However, in a {\em driven}
(sheared) system
inhomogeneities are weaker than for {\em free} cooling systems where
clusters grow and grow without being disturbed by the action of the
energy input. For this reason,
the magnitude of the differences between the Enskog predictions
and MD simulations observed here are in general smaller than those
previously reported in the undriven homogeneous case \cite{DHGD02,LBD02}
for the same values of
density and coefficients of restitution.
In this context,
it could be said that the range of densities and dissipation
for which the Enskog kinetic theory applies depends rather
strongly on the state of the system, and cannot be given as a general rule.
In the USF problem considered here, it seems that the Enskog equation
can be taken as a reliable model to get the dependence of the
rheological properties on the mechanical parameters of the mixture.

Nevertheless,
the (steady) USF problem is inherently non-Newtonian \cite{AL03b,SGD04}.
Due to the coupling
between the reduced shear rate (measured through the reduced temperature
$\theta$) and dissipation, large gradients can occur as the system
becomes more inelastic. This implies that the shear viscosity
coefficient in USF cannot be deduced from the usual Navier-Stokes
or Newtonian hydrodynamic equations (which are linear in the gradients)
\cite{SGD04}. The comparison carried out in Fig.\ \ref{fig6}
in the case of hard spheres shows that the Navier-Stokes predictions
obtained from a recent Chapman-Enskog solution \cite{MG03}
deviate from the ESMC results obtained for USF as the coefficient
of restitution decreases.
However, the agreement found between theory
and Monte Carlo simulations extends beyond the quasielastic limit
(say for instance, for $\alpha \simeq 0.97$),
except for the temperature ratio.
In this sense,
it is possible that the failure of the Navier-Stokes approximation
to describe non-Newtonian properties could be in part mitigated
by the introduction of appropriate reduced quantities, such as
those defined by Eqs.\ (\ref{3.8})--(\ref{3.10}).

Given that the dynamical velocity correlations measured in
previous works \cite{SM01,SPM01,PTNE01}
become important as the fluid
density increases, some authors \cite{SM01,BM04} conclude that
the Enskog equation can be insufficient to compute average properties
of the fluid. This conclusion contrasts with the results
reported in this paper
where, at least for the problem studied here,
velocity correlations do not seem to play a relevant role and
the Enskog equation accurately predicts the rheological properties
of the system.
It is possible that for more complex situations,
velocity correlations become important even before the system develops
significant spatial correlations and the Enskog theory
(which only takes into account spatial correlations) does not
provide reliable predictions. In this case, new kinetic theories
incorporating the effect of velocity correlations  are needed
to describe dense granular fluids at finite dissipation.

\acknowledgments

Partial support of the Ministerio de Ciencia y Tecnolog\'{\i}a (Spain) through
Grant No. ESP2003-02859 (partially financed by FEDER funds)
in the case of J.M.M. and FIS2004-01399 (partially
financed by FEDER funds) in the case of V.G. is acknowledged.
M.A. and S.L. acknowledge partial supports from the AvH Foundation and JNCASR,
and S.L. acknowledges partial support from the DFG.

\end{document}